\ifx\compilefullpaper\undefined
\documentclass[10pt, twocolumn, aps, nofootinbib, longbibliography, nobibnotes, superscriptaddress]{revtex4-1} 
\usepackage{amsfonts, amssymb, amsmath, amsthm}
\usepackage{siunitx} 
\usepackage{latexsym}
\usepackage[tracking=smallcaps]{microtype}	
\usepackage{url}
\usepackage{color}
\definecolor{DarkGray}{rgb}{0.1,0.1,0.5}
\definecolor{curgray}{rgb}{0.5,0.5,0.5}
\definecolor{curblue}{rgb}{0.04,0.11,0.64}
\definecolor{curpurple}{rgb}{0.65,0.16,0.58}
\definecolor{curorange}{rgb}{1,0.32,0}

\PassOptionsToPackage{hyphens}{url}

\usepackage[colorlinks=true,breaklinks=true, linkcolor=DarkGray,citecolor=DarkGray,urlcolor=DarkGray]{hyperref}	

\usepackage{graphicx}
\usepackage{subfigure}

\newcommand{\ket}[1]{{|#1\rangle}}

\DeclareMathOperator{\Ex}{\operatorname{E}}

\newcommand{\BQP}{\ensuremath{\mathsf{BQP}}}
\newcommand{\BPP}{\ensuremath{\mathsf{BPP}}}

\DeclareMathOperator{\CNOT}{\operatorname{CNOT}}
\DeclareMathOperator{\SWAP}{\operatorname{SWAP}}

\def\CZ {C\!Z}	

\newcounter{sprows}

\newlength{\spheight}
\newlength{\spraise}

\newlength{\commentslength}

\newcommand{\rem}[1]{}

\newfont{\subsubsecfnt}{ptmri8t at 11pt}
\renewcommand{\subparagraph}[1]{\smallskip{\subsubsecfnt #1.}}

\newcommand{\eqnref}[1]{\hyperref[#1]{{(\ref*{#1})}}}
\newcommand{\thmref}[1]{\hyperref[#1]{{Theorem~\ref*{#1}}}}
\newcommand{\lemref}[1]{\hyperref[#1]{{Lemma~\ref*{#1}}}}
\newcommand{\corref}[1]{\hyperref[#1]{{Corollary~\ref*{#1}}}}
\newcommand{\defref}[1]{\hyperref[#1]{{Definition~\ref*{#1}}}}
\newcommand{\secref}[1]{\hyperref[#1]{{Sec.~\ref*{#1}}}}
\newcommand{\figref}[1]{\hyperref[#1]{{Fig.~\ref*{#1}}}}  
\newcommand{\figureref}[1]{\hyperref[#1]{{Figure~\ref*{#1}}}}  
\newcommand{\tabref}[1]{\hyperref[#1]{{Table~\ref*{#1}}}}
\newcommand{\remref}[1]{\hyperref[#1]{{Remark~\ref*{#1}}}}
\newcommand{\appref}[1]{\hyperref[#1]{{Appendix~\ref*{#1}}}}
\newcommand{\claimref}[1]{\hyperref[#1]{{Claim~\ref*{#1}}}}
\newcommand{\factref}[1]{\hyperref[#1]{{Fact~\ref*{#1}}}}
\newcommand{\propref}[1]{\hyperref[#1]{{Proposition~\ref*{#1}}}}
\newcommand{\exampleref}[1]{\hyperref[#1]{{Example~\ref*{#1}}}}
\newcommand{\conjref}[1]{\hyperref[#1]{{Conjecture~\ref*{#1}}}}

\allowdisplaybreaks[1]

\def\COLOR{}
\ifdefined\COLOR

\else

\fi

\definecolor{Cayenne}{rgb}{0.5,0,0}
\definecolor{Midnight}{rgb}{0,0,0.5}
\definecolor{Plum}{rgb}{0.5,0,0.5}
\definecolor{Teal}{rgb}{0,0.5,0.5}
\definecolor{Clover}{rgb}{0,0.5,0}
\definecolor{Maroon}{rgb}{0.5,0,0.25}
\definecolor{Ocean}{rgb}{0,0.25,0.5}
\definecolor{Tangerine}{rgb}{1,0.5,0}
\definecolor{Strawberry}{rgb}{1,0,0.5}
\definecolor{Fern}{rgb}{0.25,0.5,0}
\definecolor{Aqua}{rgb}{0,0.5,1}
\definecolor{Moss}{rgb}{0,0.5,0.25}
\definecolor{Mocha}{rgb}{0.5,0.25,0}
\definecolor{Lemon}{rgb}{1,1,0}
\definecolor{Asparagus}{rgb}{0.5,0.5,0}
\definecolor{Grape}{rgb}{0.5,0,1}
\definecolor{Iron}{rgb}{.3,.3,.3}
\definecolor{Steel}{rgb}{.4,.4,.4}
\definecolor{Purple}{rgb}{.5,0,.5}

\def\llbracket{{[\![}}
\def\rrbracket{{]\!]}}
\usepackage[utf8]{inputenc}
\usepackage[table]{xcolor}\usepackage{soul}
\usepackage{afterpage} 

\usepackage{makecell} 

\usepackage[only,llbracket,rrbracket]{stmaryrd}

\usepackage{enumitem} 
\usepackage{color}
\definecolor{cardinal}{rgb}{0.827, 0, 0}

\begin{document}

\fi

\vfuzz2pt 

\title{
Fault-tolerant quantum computation with a neutral atom processor
}

\affiliation{Microsoft Quantum} 
\affiliation{Atom Computing}

\author{Ben W. Reichardt}
\affiliation{Microsoft Quantum}
\affiliation{Department of Electrical and Computer Engineering, University of Southern California}

\author{Adam Paetznick}
\affiliation{Microsoft Quantum}

\author{David Aasen}
\affiliation{Microsoft Quantum}

\author{Ivan Basov}
\affiliation{Microsoft Quantum}

\author{Juan M.~Bello-Rivas}
\affiliation{Microsoft Quantum}

\author{\\ Parsa Bonderson}
\affiliation{Microsoft Quantum}

\author{Rui Chao}
\affiliation{Microsoft Quantum}

\author{Wim van Dam}
\affiliation{Microsoft Quantum}

\author{Matthew B.~Hastings}
\affiliation{Microsoft Quantum}

\author{Ryan V. Mishmash}
\affiliation{Microsoft Quantum}

\author{Andres Paz}
\affiliation{Microsoft Quantum}

\author{Marcus P.~da Silva}
\affiliation{Microsoft Quantum}

\author{\\ Aarthi Sundaram}
\affiliation{Microsoft Quantum}

\author{Krysta M.~Svore}
\affiliation{Microsoft Quantum}

\author{Alexander Vaschillo}
\affiliation{Microsoft Quantum}

\author{Zhenghan Wang}
\affiliation{Microsoft Quantum}

\author{Matt Zanner}
\affiliation{Microsoft Quantum}

\author{\\ William~B.~Cairncross}
\affiliation{Atom Computing}

\author{Cheng-An~Chen}
\affiliation{Atom Computing}

\author{Daniel~Crow}
\affiliation{Atom Computing}

\author{Hyosub~Kim}
\affiliation{Atom Computing}

\author{Jonathan~M.~Kindem}
\affiliation{Atom Computing}

\author{Jonathan~King}
\affiliation{Atom Computing}

\author{Michael~McDonald}
\affiliation{Atom Computing}

\author{Matthew~A.~Norcia}
\affiliation{Atom Computing}

\author{Albert~Ryou}
\affiliation{Atom Computing}

\author{Mark~Stone}
\affiliation{Atom Computing}

\author{Laura~Wadleigh}
\affiliation{Atom Computing}

\author{\\Katrina~Barnes}
\affiliation{Atom Computing}

\author{Peter~Battaglino}
\affiliation{Atom Computing}

\author{Thomas~C.~Bohdanowicz}
\affiliation{Atom Computing}

\author{Graham~Booth}
\affiliation{Atom Computing}

\author{Andrew~Brown}
\affiliation{Atom Computing}

\author{Mark~O.~Brown}
\affiliation{Atom Computing}

\author{Kayleigh~Cassella}
\affiliation{Atom Computing}

\author{Robin~Coxe}
\affiliation{Atom Computing}

\author{Jeffrey~M.~Epstein}
\affiliation{Atom Computing}

\author{Max~Feldkamp}
\affiliation{Atom Computing}

\author{Christopher~Griger}
\affiliation{Atom Computing}

\author{Eli~Halperin}
\affiliation{Atom Computing}

\author{Andre~Heinz}
\affiliation{Atom Computing}

\author{Frederic~Hummel}
\affiliation{Atom Computing}

\author{Matthew~Jaffe}
\affiliation{Atom Computing}

\author{Antonia~M.~W.~Jones}
\affiliation{Atom Computing}

\author{Eliot~Kapit}
\affiliation{Atom Computing}
\affiliation{Department of Physics, Colorado School of Mines}

\author{Krish~Kotru}
\affiliation{Atom Computing}

\author{Joseph~Lauigan}
\affiliation{Atom Computing}

\author{Ming~Li}
\affiliation{Atom Computing}

\author{Jan~Marjanovic}
\affiliation{Atom Computing}

\author{Eli~Megidish}
\affiliation{Atom Computing}

\author{Matthew~Meredith}
\affiliation{Atom Computing}

\author{Ryan~Morshead}
\affiliation{Atom Computing}

\author{Juan~A.~Muniz}
\affiliation{Atom Computing}

\author{Sandeep~Narayanaswami}
\affiliation{Atom Computing}

\author{Ciro~Nishiguchi}
\affiliation{Atom Computing}

\author{Timothy~Paule}
\affiliation{Atom Computing}

\author{Kelly~A.~Pawlak}
\affiliation{Atom Computing}

\author{Kristen~L.~Pudenz}
\affiliation{Atom Computing}

\author{David~Rodr\'iguez~P\'erez}
\affiliation{Atom Computing}

\author{Jon~Simon}
\affiliation{Atom Computing}
\affiliation{Department of Physics and Department of Applied Physics, Stanford University}

\author{Aaron~Smull}
\affiliation{Atom Computing}

\author{Daniel~Stack}
\affiliation{Atom Computing}

\author{Miroslav~Urbanek}
\affiliation{Atom Computing}

\author{Ren\'e~J.~M.~van de Veerdonk}
\affiliation{Atom Computing}

\author{Zachary~Vendeiro}
\affiliation{Atom Computing}

\author{Robert~T.~Weverka}
\affiliation{Atom Computing}

\author{Thomas~Wilkason}
\affiliation{Atom Computing}

\author{Tsung-Yao~Wu}
\affiliation{Atom Computing}

\author{Xin~Xie}
\affiliation{Atom Computing}

\author{Evan~Zalys-Geller}
\affiliation{Atom Computing}

\author{Xiaogang~Zhang}
\affiliation{Atom Computing}

\author{Benjamin~J.~Bloom}
\affiliation{Atom Computing}

\begin{abstract}

Quantum computing experiments are transitioning from running on physical qubits to using encoded, logical qubits.  Fault-tolerant computation can identify and correct errors, 
and has the potential to enable the dramatically reduced logical error rates required for valuable algorithms.  However, it requires flexible control of high-fidelity operations performed on large numbers of qubits.  We demonstrate fault-tolerant quantum computation on a quantum processor with $256$ qubits, each an individual neutral Ytterbium atom.  
The operations 
are designed so that key error sources convert to atom loss, which can be detected by imaging.  
Full connectivity is enabled by atom movement.  
We demonstrate the entanglement of $24$ logical qubits encoded into $48$ atoms, at once catching errors and {correcting} for, on average 1.8, lost atoms.  We also implement the Bernstein-Vazirani algorithm with up to $28$ logical qubits encoded into $112$ atoms, showing better-than-physical error rates.  In both cases, 
``erasure conversion," changing errors into a form that can be detected independently from qubit state, 
improves circuit performance.  These results begin to clear a path for achieving scientific quantum advantage with a programmable neutral atom quantum processor.  
\end{abstract}

\maketitle




Quantum computers are potentially powerful, but are challenging to control precisely.  
State-of-the-art two-qubit gate error rates 
in some flagship quantum processors range 
from $0.1\%$ for a trapped-ion device~\cite{Moses23quantinuum} to an average of $0.4\%$ for a superconducting qubit device~\cite{google24surfacecode}.  
Moderate-size quantum algorithms require much lower error rates, 
below $10^{-6}$~\cite{Beverland22quantumadvantage}. 
This precision can likely only be achieved 
using quantum error-correcting codes to store and fault-tolerantly operate on ``logical" qubits~\cite{AliferisGottesmanPreskill05}.  
%
%
Here, we advance the shift from physical to logical qubits, by demonstrating fault-tolerant, encoded computations on logical qubits.  The encoded circuits outperform unencoded circuits on physical qubits.  

\begin{figure}
\includegraphics[scale=.17]{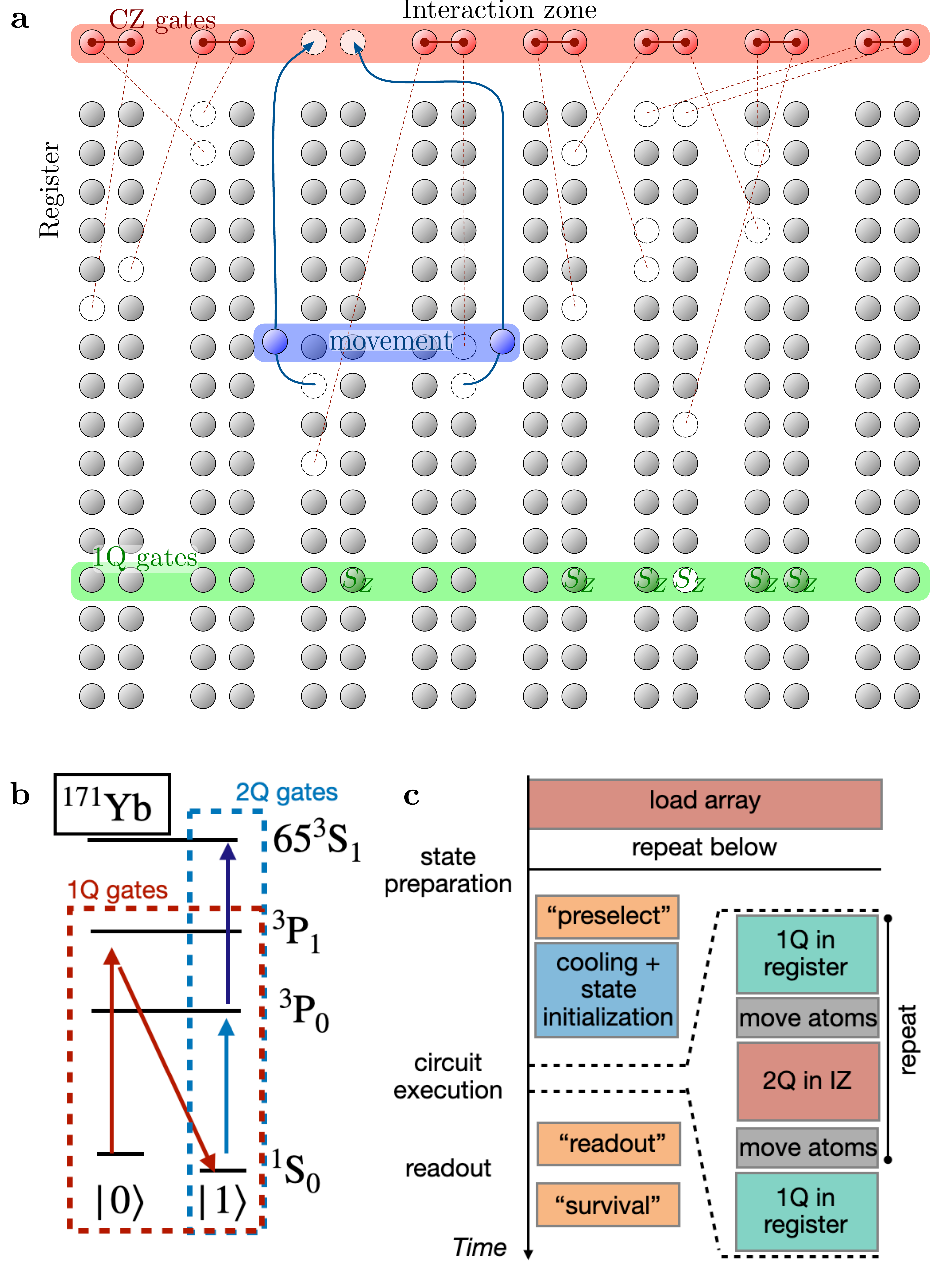}
\caption{
Experimental architecture.  {\bf a}, {\bf b}, 
$^{171}$Yb atoms, each giving a physical qubit in the ground-state nuclear spin, can occupy tweezer traps within either a register array or an interaction zone (IZ).  Up to 8 pairs of atoms interact at a time in the IZ, driven by two-qubit (2Q) gate lasers to implement controlled-phase (CZ) gates.  These lasers perform coherent sequential excitation to a metastable clock state ($^{3}$P${_0}$) and the Rydberg state (65$^{3}$S${_1}$). Arbitrary single-qubit (1Q) operations can be applied in parallel to the atoms within a register row using Raman transitions detuned from the $^{3}$P${_1}$ manifold.  Moving atoms between the zones enables arbitrary 2Q gate connectivity.  {\bf c}, Circuit-running sequence.  Dissipative operations (left column) are performed with atoms held in a cavity-enhanced optical lattice.  Three images 
(orange boxes) allow us to determine initial occupancy of register sites, as well as the final state and presence of atoms.  Coherent operations (right column) are performed with atoms held in tweezers.  
}
\label{f:system}
\end{figure}

\begin{figure}
\includegraphics[width=\columnwidth]{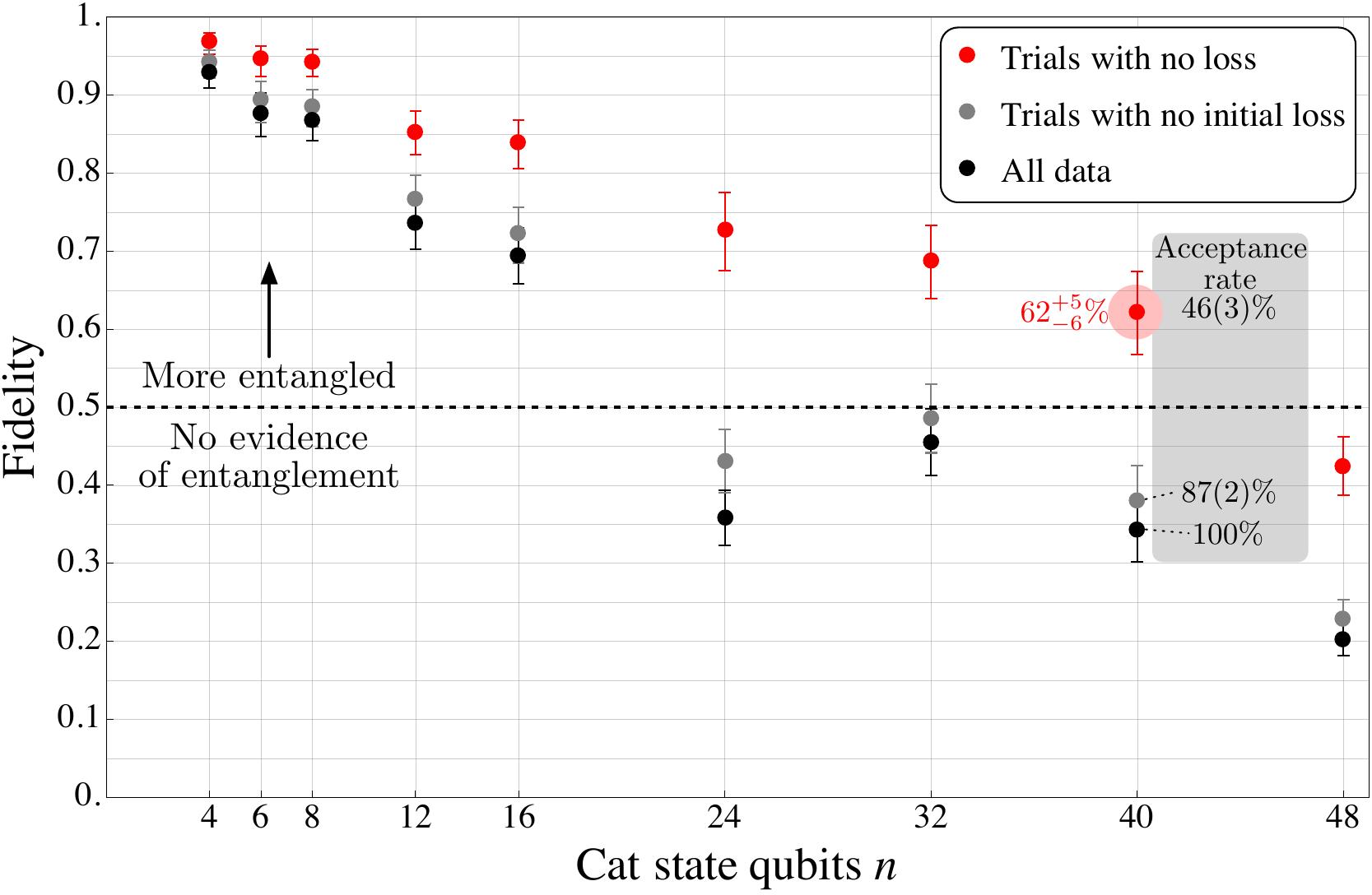}
\caption{
Fidelity of prepared cat states versus qubit number~$n$.  Red points give fidelities for trials with no lost atoms.  
Cat states with up to $40$ qubits exceed the entanglement threshold.  Gray points include trials in which all atoms are present initially but atoms lost during the circuit are replaced by the maximally mixed state, and black points include all trials.  
}
\label{f:ghz}
\end{figure}

\begin{figure}
{\raisebox{0cm}{\includegraphics[scale=.29]{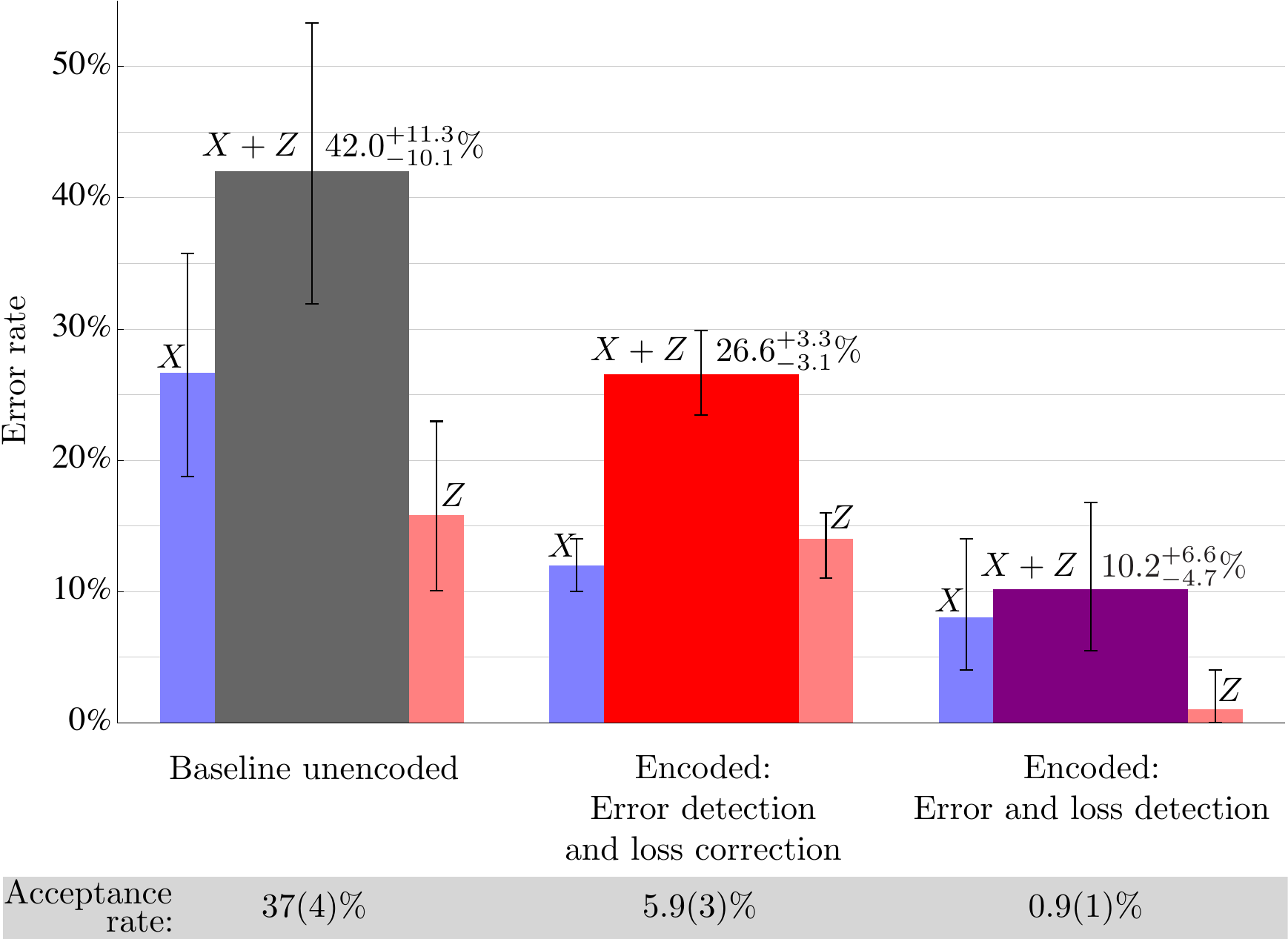}}}
\caption{
Encoded $24$-qubit cat state results.  Shown is the error for measurements in the $X$ basis, the error for measurements in the $Z$ basis, and their sum, for three cases.  
The baseline is an unencoded $24$-qubit cat state, based on trials with no lost atoms.  
For the encoded cat state data, in one case we reject trials with a detected error, but try to correct for qubits lost during the circuit.  In the other case, we reject trials with any lost qubit or detected error.   
}
\label{f:cat24summary}
\end{figure}

\begin{figure}
{\raisebox{0cm}{\includegraphics[scale=.3]{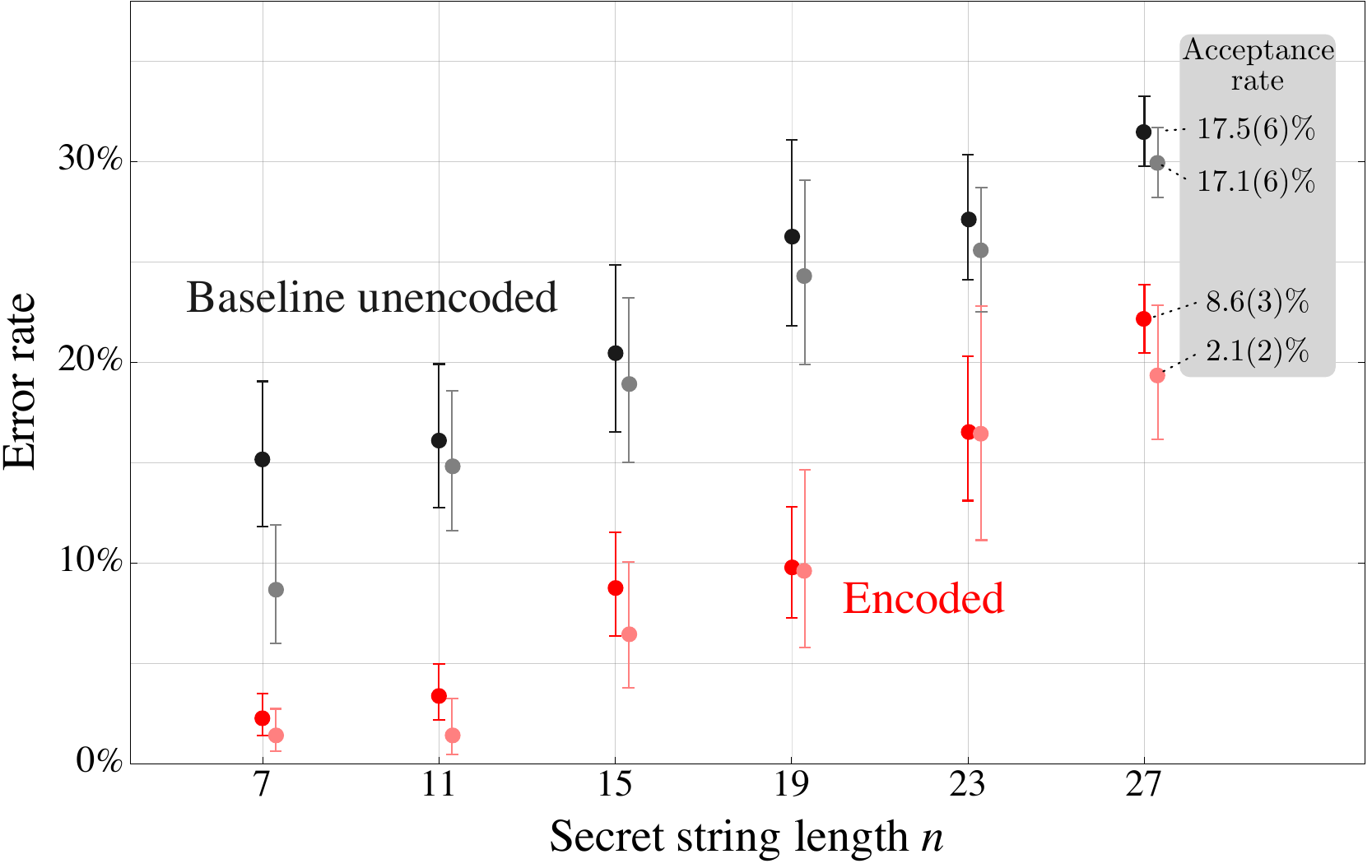}}}
\caption{
Results for the Bernstein-Vazirani algorithm with an $n$-bit secret string.  The string is always $1^n$, as the corresponding oracle is the most difficult to implement.  
The success probability is the probability of measuring $1^n$, conditioned on acceptance.  
For the unencoded algorithm (black), we accept when the $n$ measured atoms are present.  For the encoded algorithm (red), we accept when all of the $n$ measured code blocks are decodable; in particular, it can detect faults and correct for lost atoms.  
The lighter points are based only on those trials with no initial atom loss.  
}
\label{f:bvsummary}
\end{figure}

Our results are achieved using a neutral atom quantum processor with $256$ qubits, Ytterbium atoms (\figref{f:system}), designed to enable loss conversion and detection for improved logical performance.  
Neutral atom devices have been mainly analog, 
but recent progress has enabled 
digital, gate-based computation~\cite{evered2023high,scholl2023erasure,ma2023high,radnaev2024,Finkelstein2024,muniz2024}.
The processor here has low gate error, and 
atom movement enables all-to-all connectivity for flexible computation and logical operation design.  
The physical two-qubit gates are designed so that key error sources convert to atom loss, which imaging has been designed to detect.  With unencoded circuits, data with lost atoms can be discarded, and using this selection we prepare an entangled $40$-qubit ``cat" state (\figref{f:ghz}).  Loss conversion is even more advantageous operating on encoded data, though; 
a distance-two code can detect errors and correct for lost atoms.  
%
We demonstrate the creation and entanglement of logical qubits, and logical computation.  
We prepare a cat state on $24$ logical qubits, encoded in $48$ atoms (\figref{f:cat24summary}).  
We then demonstrate a quantum algorithm on logical qubits.  
We fault-tolerantly run the Bernstein-Vazirani algorithm~\cite{BernsteinVazirani97complexity,cemm:qalg}, 
well known for showing the higher efficiency of quantum computers versus classical, 
on as many as $28$ logical qubits, encoded in $112$ atoms (\figref{f:bvsummary}), with better-than-physical results.  This illustrates the scale available in neutral atom processors, and their promise as reliable, fault-tolerant quantum processors.  

Creating and verifying large cat or GHZ states~\cite{greenberg1989ghz} 
is a common benchmark for hardware progress~\cite{Krenn24catstates}.
The current record is a $60$-qubit cat state, prepared with $59\%$ fidelity using superconducting qubits~\cite{Bao24catstate60}.  A $20$-qubit cat state with $54\%$ fidelity, after adjusting for measurement error, has been prepared on a neutral atom system~\cite{Omran19neutexp}.  
The $62^{+5}_{-6}\%$ fidelity of our $40$-qubit cat state is not adjusted for measurement errors.  

\tabref{f:catstateexperiments} summarizes results with encoded cat states.  
A cat state on $12$ logical qubits has been prepared with 
$99.8\%$ fidelity~\cite{ReichardtMicrosoft24tesseract}, on a trapped-ion processor~\cite{Moses23quantinuum}.  
On a neutral atom system, Bluvstein et al.~\cite{BluvsteinHarvard23neutralatoms} have prepared a four-qubit encoded cat state 
with fidelity $72(2)\%$ using error correction, ranging to $99.85^{+0.1}_{-1.0}\%$ using full error detection and with a 
lower acceptance rate.  Our $24$-qubit encoded cat state is the largest published to date.

\begin{table}
\caption{\label{f:catstateexperiments}
Experiments preparing encoded cat states.}
\setlength{\tabcolsep}{1.9pt}
\begin{tabular}{ccc@{$\quad$}c}
\Xhline{2\arrayrulewidth}
& Logical \\[-.1cm]
Reference & qubits & Code & Total error $p_X + p_Z$ \\
\hline
\cite{Hong24cat4onh2} & 4 & $\llbracket25,4,3\rrbracket$ & $0.5^{+0.4}_{-0.2}\%$ \\ 
\cite{BluvsteinHarvard23neutralatoms} & 4 & $\llbracket7,1,3\rrbracket$ & 
\scalebox{.9}{$\!\!\!\left\{\begin{array}{c} \\[.20cm] \end{array}\right.\!\!\!\!\!$}
\begin{tabular}{c}28(2)\% error correction\\ $0.15^{+1.0}_{-0.1}\%$ error detection\end{tabular}$\!\!\!$ \\
\cite{ReichardtMicrosoft24tesseract} & 12 & $\llbracket16,4,4\rrbracket$ & 
$0.18^{+0.4}_{-0.12}\%$ \\
This work & 24 & $\llbracket4,2,2\rrbracket$ & 
\scalebox{.9}{$\!\!\!\left\{\begin{array}{c} \\[1.1cm] \end{array}\right.\!\!\!\!\!\!\!$}
\begin{tabular}{c}
$26.6^{+3.3}_{-3.1}\%$ error detection \\[-.1cm] and loss correction\\
$10.2^{+6.6}_{-4.7}\%$ error and \\[-.1cm] loss detection\\
\end{tabular} \\
\Xhline{2\arrayrulewidth}
\end{tabular}
\end{table}

The Bernstein-Vazirani algorithm for finding a secret string is another benchmark for quantum processors~\cite{Lubinski24qedcbenchmarks}.
It has been implemented with unencoded qubits on superconducting processors~\cite{Roy20bernsteinvazirani, DahlhauserHumble20bernsteinvazirani, PokharelLidar23bernsteinvazirani}, ion trap processors~\cite{FallekBrown16bernsteinvazirani, DebnathMonroe16iontrap, Wright19iontrap11benchmark}, and both processor types~\cite{Linke17bernsteinvaziraniiontrapsuperconducting, Murali19bernsteinvazirani, BlinovWuMonroe21bernsteinvaziraniiontrapsuperconducting, Lubinski21algorithmbenchmarks}, for various lengths of the secret string, up to length $26$ for superconducting qubits~\cite{PokharelLidar23bernsteinvazirani} and up to length $20$ for ion traps~\cite{Lubinski21algorithmbenchmarks}.  It is difficult to compare these results fairly.  Some test identification of the secret $s = 1^n$, since it is hardest to implement; the corresponding oracle requires the most two-qubit gates.  Others test against random strings~$s$, 
and this is much easier, 
comparable to finding $s = 1^{n/2}$.  The success probabilities also vary considerably, e.g., from $0.2\%$ 
for $s = 1^{26}$~\cite{PokharelLidar23bernsteinvazirani}, to $70\%$ for random $s$ with $n = 20$~\cite{Lubinski21algorithmbenchmarks}.  Our fault-tolerant, encoded implementation of the Bernstein-Vazirani algorithm is the first to date, and has the largest length, $n = 27$.  On $s = 1^{27}$, the $91.4(3)\%$ success probability for the encoded circuit beats the unencoded algorithm.

\section*{
Neutral atom quantum processor}

The quantum processor used in this work is based on reconfigurable arrays of neutral $^{171}$Yb atoms, depicted in \figref{f:system}a,b,  with the qubits encoded in the ground-nuclear spin states ($^1$S$_0$, $m_F = \pm 1/2$).  Aspects of the system have been described in previous publications~\cite{barnes2022assembly, norcia2023midcircuit, norcia2024iterative, muniz2024}, and further details can be found in the methods section.  
By moving atoms between arrays of static optical tweezer traps using mobile tweezers, this platform enables arbitrary connectivity between large numbers of qubits.  

Figure~\ref{f:system}c shows a typical sequence for circuit execution.  State preparation consists of iterative loading of the relevant sites of the register array, as described in~\cite{norcia2024iterative}, followed by an image to determine successful filling of sites, cooling, and initialization of the $\ket{0}$ state through optical pumping.  State-resolved, low-loss readout as well as cooling and state initialization are performed by transferring atoms from the tweezer arrays into traps formed by cavity-enhanced optical lattices, which allows for state and atom detection with error and loss rates at the part-per-thousand level (see methods).  
All other operations are peformed with the atoms held in the tweezer arrays. For this work, up to 256 atoms are available for computation.  

During circuit execution, single-qubit operations and qubit storage are performed in a register array.  Two-qubit (2Q) gates are performed on up to eight qubit pairs in parallel within a separate interaction zone (IZ), with connectivity determined by the placement of atoms within the IZ.  2Q gates are performed using state-selective sequential two-photon excitation to a high-lying Rydberg state in a blockaded regime, resulting in controlled-phase (CZ) gates between pairs of qubits.  Importantly, key sources of 2Q gate errors are converted to loss, 
which can be detected independently from the qubit state. 
Population errors in the sequential excitation used in the gates lead either directly to loss, or to population of metastable states that are not trapped in register traps.   Unlike most systems based on individually controlled neutral atoms, our ability to perform nondestructive state-selective imaging \cite{norcia2023midcircuit} allows us to perform a pair of images that unambiguously distinguish between three possibilities for the atomic state: $\ket{0}$, $\ket{1}$, and lost.

\tabref{tab:rb} presents a summary of important error rates, with benchmark protocols described in the methods.  

\begin{table}
\caption{Randomized benchmark (RB) infidelities, conditioned on an atom (1Q) or pair of atoms (2Q) staying in the qubit subspace, and probability of losing one or both atoms.  
} \label{tab:rb}
\centering
\begin{tabular}{ccc}
\Xhline{2\arrayrulewidth}
Benchmark protocol & Infidelity & Loss + Leakage\\
\hline
2Q Static IRB & 0.56(15)\% & 0.27(6)\%\\
2Q Echoed RB with moves & 0.39(10)\% & 0.75(10)\%\\
2Q Echoed RB without moves & 0.35(5)\% & 0.24(6)\%\\
\Xhline{2\arrayrulewidth}
\end{tabular}
\end{table}

\section*{Cat state generation}


We prepare non-encoded cat states using modified versions of the circuits from \cite{cruz2019efficient}.  
We generate the ``antiferromagnetic" cat state $\frac{1}{\sqrt{2}}(\ket{0101\ldots} + \ket{1010\ldots})$, instead of the ``ferromagnetic" state $\frac{1}{\sqrt{2}}(\ket{0^n} + \ket{1^n})$, since it is less sensitive to correlated phase errors.  
%
%
%
%
Figure~\ref{f:ghz} plots the fidelity to the ideal 
cat state, computed as $1 - \tfrac12(Z$ basis error rate$) - ($average error rate over $n$ bases orthogonal to~$z)$~\cite{guhne2007toolbox}.  A fidelity above 50\% certifies $n$-partite entanglement~\cite{G_hne_2010}.  
Restricting the analysis to only those trials where all atoms survived, the cat states exceed the entanglement threshold up to $n = 40$.  To our knowledge, this is the largest cat state demonstrated with neutral atoms.  

We can also estimate the fidelity including all trials, or all trials with no initial loss, in the analysis generously treating lost qubits as replaced by the maximally mixed state.  This gives a higher acceptance rate, but lower fidelity, showing the advantage of loss detection.  



\section*{Encoded $24$-qubit cat state}

\begin{figure}
{\raisebox{0cm}{\includegraphics[scale=.6]{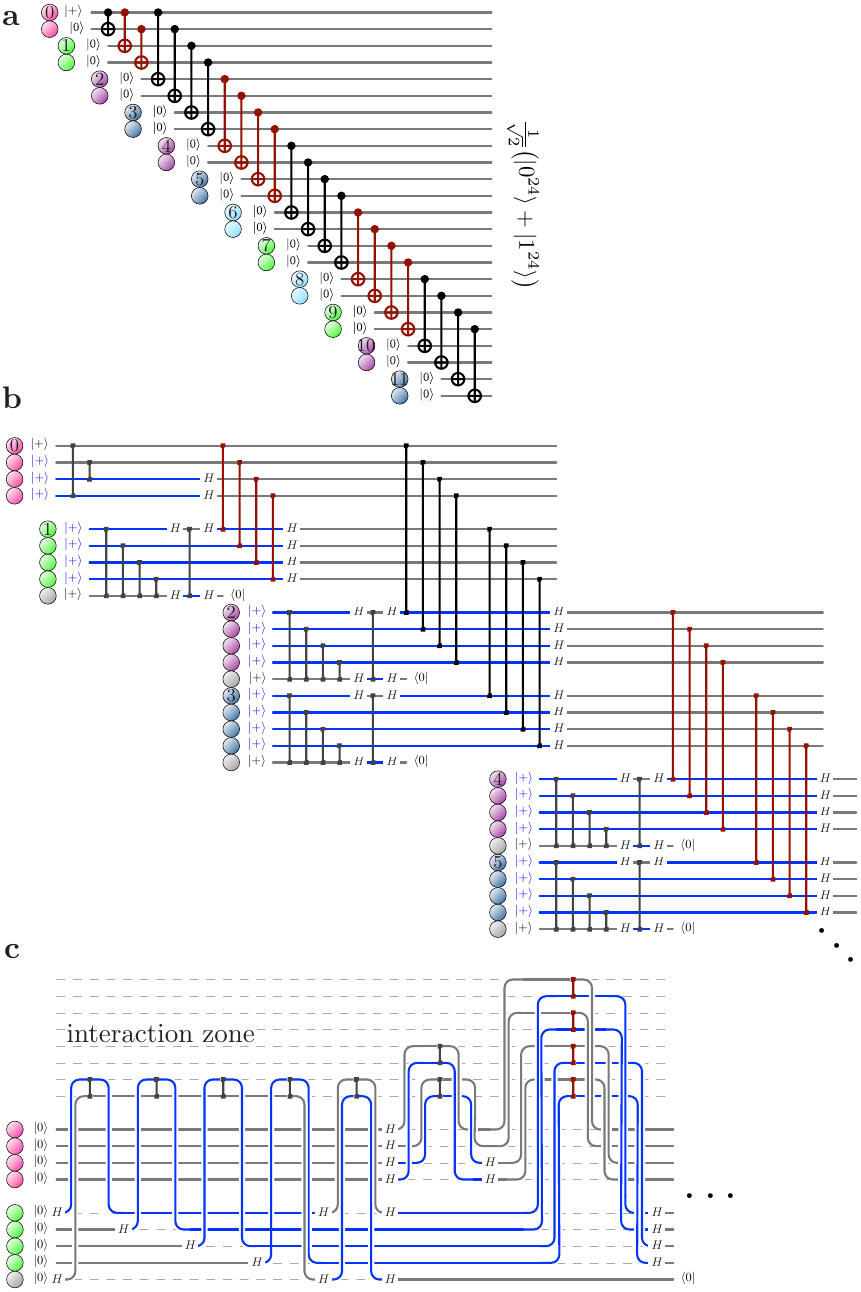}}}
\caption{
{\bf a},
Logical circuit to prepare a $24$-qubit cat state.  
The qubits are arranged in $12$ pairs that will each be encoded into a $\llbracket 4,2,2\rrbracket$ code block.  
{\bf b},
The encoded circuit 
involves $59$ atoms 
and $101$ physical CZ gates.  
All but the first code block use extra ancilla qubits, shown in gray, to prepare encoded $\ket{00}$ fault tolerantly; each ancilla should be measured to be $0$.  In the 
diagram, a blue wire is a qubit in the Hadamard basis.  
Not shown are the final transversal $X$ or $Z$ measurements.  
{\bf c},~Schedule for the atom movements and gates.  
}
\label{f:cat24compiledcircuit}
\end{figure}

We prepare a 
cat state on $24$ logical qubits, the largest published to date. 
 The logical qubits are encoded with the $\llbracket4,2,2\rrbracket$ code 
into $48$ physical qubits.  
Starting with a logical state-preparation 
circuit, 
we 
encode the circuit, 
then 
schedule the gates and movements (\figref{f:cat24compiledcircuit}). 

Our experiments, like \cite{Hong24cat4onh2, ReichardtMicrosoft24tesseract}, measure $X$ basis and $Z$ basis error rates, $p_X$ and $p_Z$.  
In the unencoded case, 
the fidelity to the ideal cat state is at least $1 - p_X - p_Z$.  
%
We find that for the encoded cat state, using error detection and loss correction, $p_X + p_Z = 26.6^{+3.3}_{-3.1}\%$, while for the unencoded baseline $p_X + p_Z = 42.0^{+11.3}_{-10.1}\%$.  
The results are summarized in \figref{f:cat24summary}.  
The baseline error bars are large because we report the best single run (different from in \figref{f:ghz}), which disadvantages the encoded cases.  
The encoding is beneficial, 
but affects $p_X$ and $p_Z$ differently.  

How useful is loss conversion, loss measurement, and loss correction for this encoded state?  The average loading loss rate was $1.04(2)\%$, and $57.2(6)\%$ of trials had no initially lost atoms.  
Among trials without initial losses, those that decoded correctly had an average of 2.0 and 1.6 circuit losses, for $X$ and $Z$ basis measurements, respectively.   The code successfully corrects for these lost qubits.  Still, circuit loss is a major cause of logical errors.  Those shots that decoded to a logical error had respective averages of 2.4 and 2.7 losses.  In the $Z$ basis, there are only two valid decodings, $0^{24}$ and $1^{24}$, and we can usually identify the most likely cause of a logical error.  For example, a decoding of $1111(0111)^5$ indicates that most probably the third block had a logical $XI$ error that propagated to the subsequent code blocks.  The frequencies of the most likely causes of the logical error are: $26\%$ two circuit losses, $54\%$ one loss and one $X$ error, $9\%$ two $X$ errors, $11\%$ three or more losses or errors.  

\begin{figure}
{\raisebox{0cm}{\includegraphics[scale=.3]{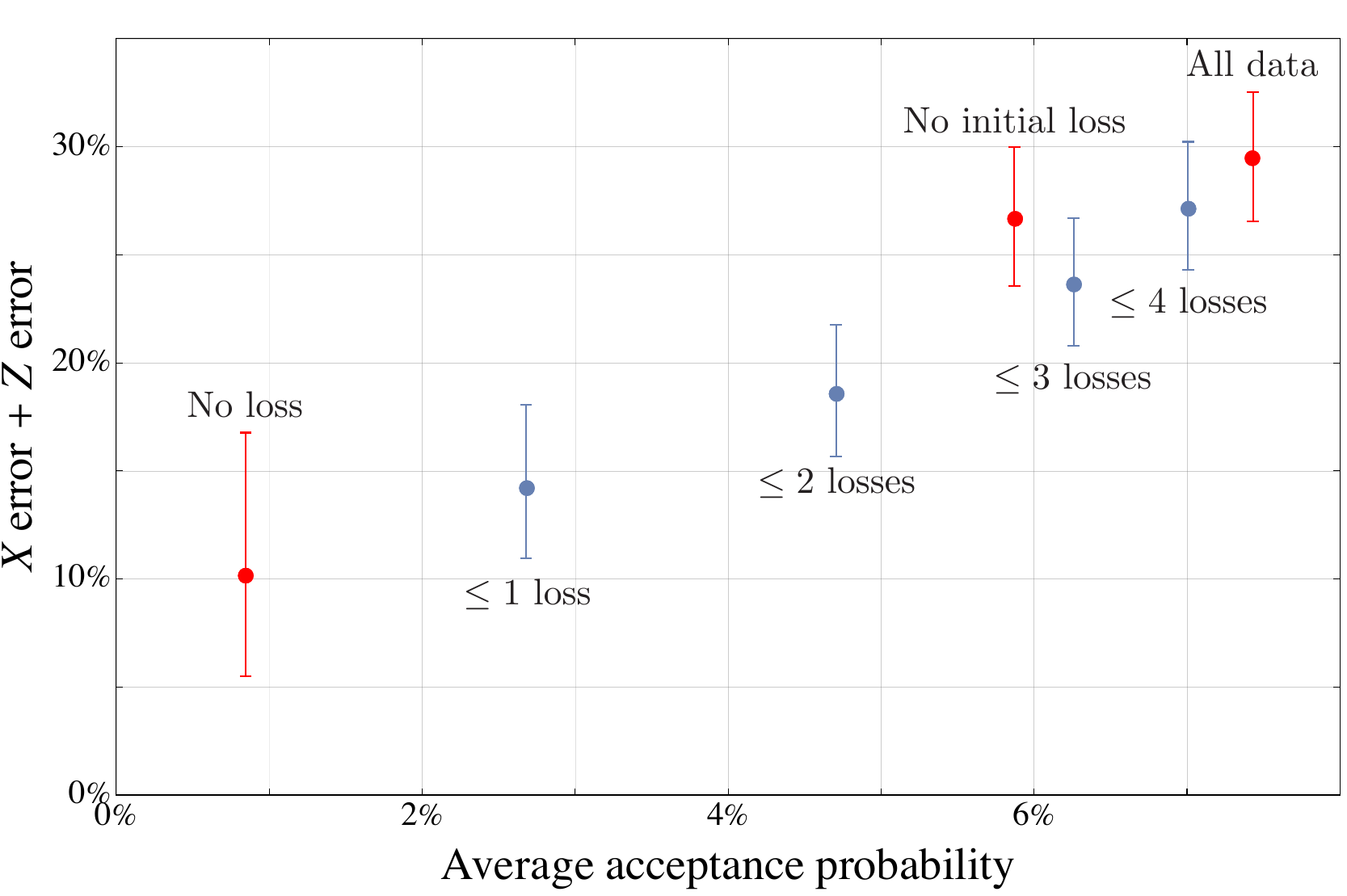}}}
\caption{
Encoded $24$-qubit cat state data, selecting only trials with specified loss criteria.  
Trying to correct more loss generally leads to a higher acceptance rate, but also a higher error rate.  
In each trial, an atom may be missing before the circuit begins 
or it may be lost during the circuit.  
The red points show the total $X$ basis plus $Z$ basis error rate, plotted against the average of the $X$ and $Z$ acceptance rates, based on either all $\num{29760}$ trials, only those $\num{17008}$ trials with no initial loss, or only those $858$ trials with no loss at all.  
The blue points select only trials in which the total number of lost qubits is at most a threshold.  
}
\label{f:cat24parsing}
\end{figure}

\begin{figure}
\raisebox{0cm}{\includegraphics[scale=.3]{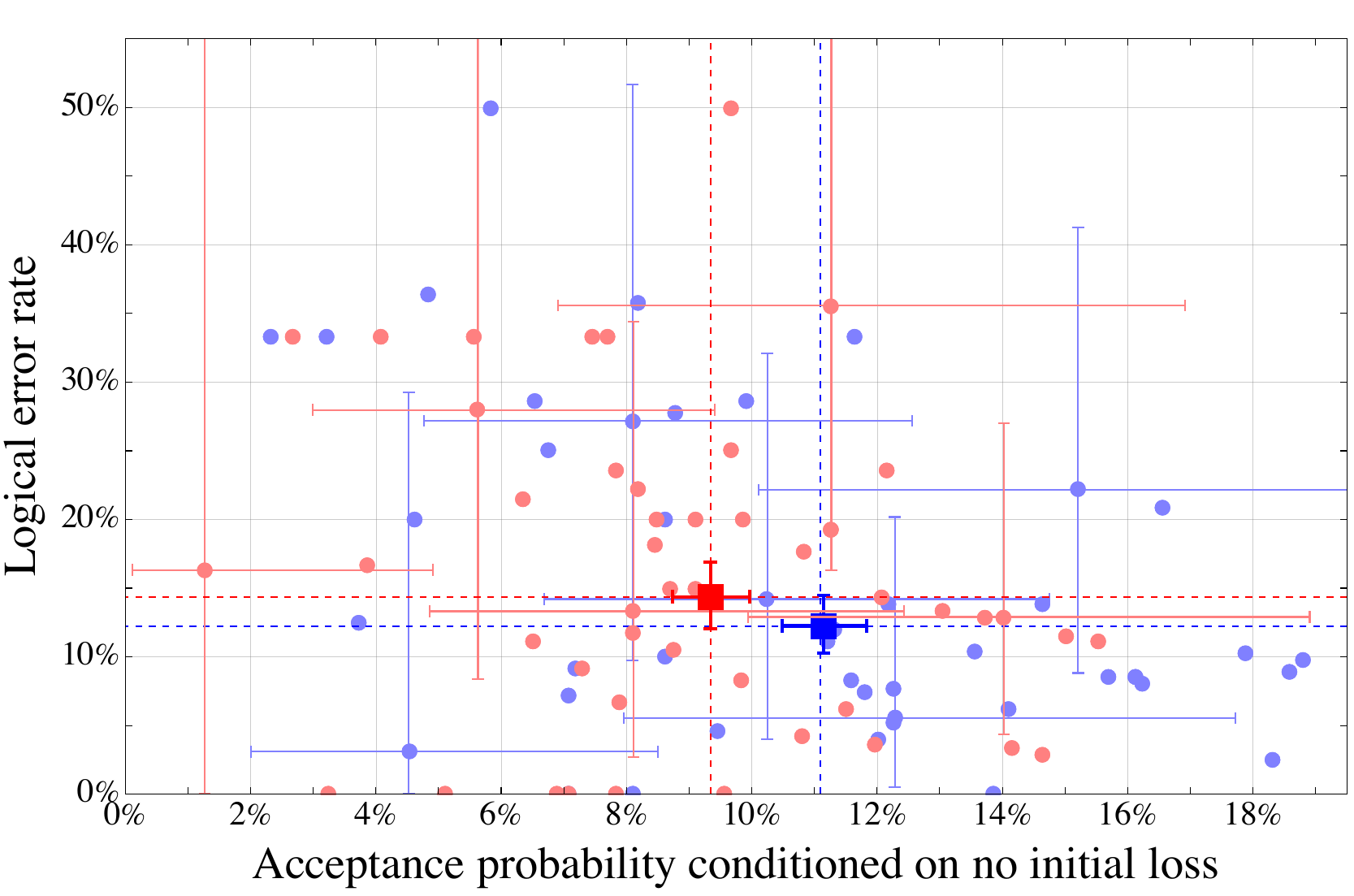}}
\caption{
Encoded $24$-qubit cat state 
runs.  
Each point gives the acceptance probability and logical error rate for a run of, on average, $183$ trials with no initial loss, blue for the $X$ basis and red for~$Z$.  
For clarity, only some error bars are shown.  The large squares are the full 
averages, 
as reported in \figref{f:cat24summary}.  The negative correlation between acceptance probability and logical error rate evidences some parameter drift between runs, despite automatic recalibrations.  
}
\label{f:cat24runs}
\end{figure}

Figure~\ref{f:cat24parsing} shows how conditioning on at most $k$ lost qubits, for $k = 0, 1, \ldots$, 
allows an interpolation 
between the different encoded results in \figref{f:cat24summary}; in particular, conditioning on no loss reduces the total error from $29.5(3.0)\%$ to $10.2^{+6.6}_{-4.7}\%$.  
Finally, \figref{f:cat24runs} shows all the encoded cat state runs, 
plotted conditioned on no initial atom loss.  The spread and correlations in the figure suggest some machine parameter drift.  

\section*{Encoded Bernstein-Vazirani algorithm}

Computing with logical qubits is the next step after encoded state preparation.  
For a secret string $s \in \{0,1\}^n$, consider a black-box oracle that on query $x \in \{0,1\}^n$ returns $s \cdot x = s_1 x_1 \oplus \cdots \oplus s_n x_n$.   
Classically, $n$ oracle queries are required to learn~$s$.  
The Bernstein-Vazirani (BV) algorithm~\cite{BernsteinVazirani97complexity,cemm:qalg} makes a single call to the corresponding quantum oracle $U_s$, $U_s \ket{b, x} = \ket{b \oplus (s \cdot x), x}$, to learn~$s$ via $H^{n+1} U_s H^{n+1} \ket{1,0^n} = \ket{1, s}$.  
This algorithm was designed to prove an oracle separation between the classical randomized and quantum polynomial-time complexity classes $\BPP$ and $\BQP$.

\begin{figure}
{\raisebox{0cm}{\includegraphics[scale=.6]{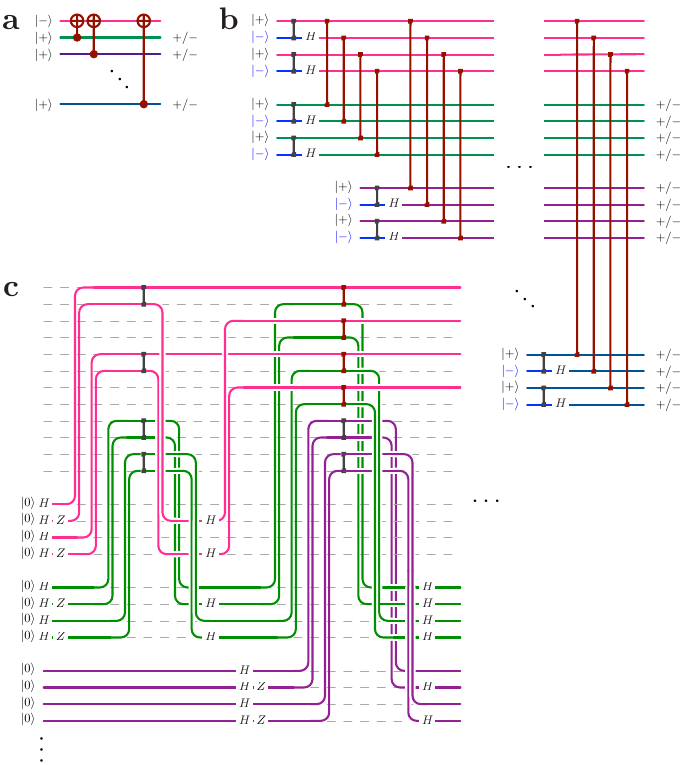}}}
\caption{
The Bernstein-Vazirani algorithm for a secret string $s \in \{0,1\}^n$ uses $n+1$ logical qubits.  
{\bf a}, 
In the most challenging case, $s = 1^n$ and there are $n$ logical CNOT gates.  
{\bf b}, Encoded into the $\llbracket 4,1,2\rrbracket$ code, 
there are $4(n+1)$ atoms and $6n+2$ 
CZ gates.  
{\bf c}, The gates are scheduled so that the first block, once initialized, stays in the interaction zone.  
As transversal CZ gates are applied into block~$j$, block~$j+1$ is initialized.  
}
\label{f:bv20atomarrangement}
\end{figure}

For $n = 7, 11, 15, \ldots, 27$, we implement the BV algorithm for the hardest secret, $s = 1^n$, using both unencoded and $\llbracket 4,1,2 \rrbracket$-encoded qubits.  
Figure~\ref{f:bv20atomarrangement} shows the encoded circuit.  
The largest case, $n = 27$, has $28$ logical qubits, $112$ atoms and $170$ 
CZ gates.   
Note that 
we encode $\ket{+}$ as $\ket{\overline{+1}} = \tfrac12 (\ket{01} + \ket{10})^{\otimes 2}$; setting the gauge qubit to $\ket 1$ lets the state be prepared fault tolerantly without verification.  
We prefer $\ket{\overline{\pm1}}$ to $\ket{\overline{\pm0}} = \tfrac12 (\ket{00} \pm \ket{11})^{\otimes 2}$ because 
its terms 
have Hamming weight two, giving a decoherence-free subspace 
to collective dephasing~\cite{PalmaSuominenEkert96dfs}.  

The unencoded algorithm accepts if no measured atom is lost, and is successful if the measured string is~$1^n$.  
The encoded algorithm tolerates losses, but rejects if any of the last $n$ code blocks is not decodable.  

Figure~\ref{f:bvsummary} summarizes the results.  
The encoded algorithms have at least a $1.4 \times$ lower error rate 
than the unencoded baseline, albeit with a lower acceptance rate. 

\begin{figure}
{\raisebox{0cm}{\includegraphics[scale=.30]{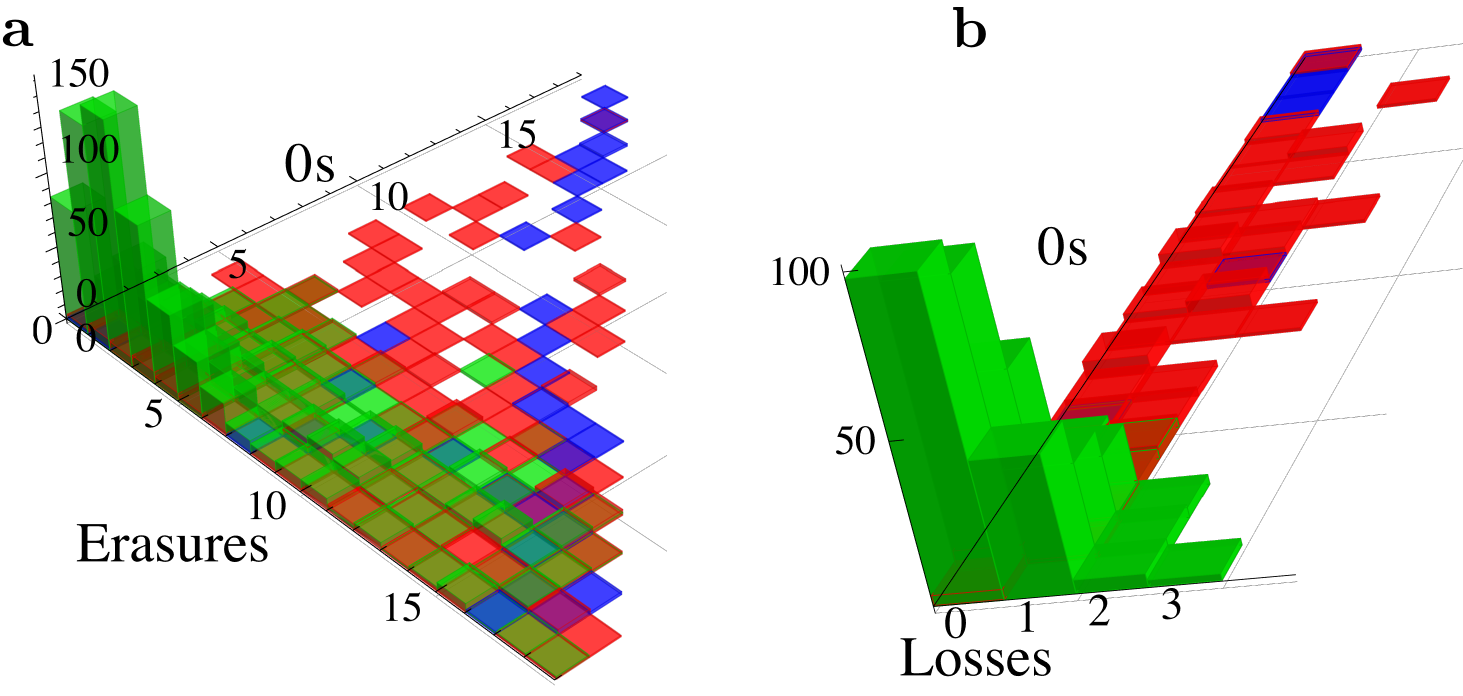}}}
\caption{
{\bf a}, Histogram of decoded values for the last $n$ code blocks, for the $n = 19$ encoded Bernstein-Vazirani algorithm.  Ideally, every block should decode to $1$, with no undecodeable blocks or $0$s.  The green bars are conditioned on the first block decoding to $\ket{-}$, the red bars on it detecting an error, and the blue bars on it decoding to $\ket{+}$, a logical error.  Note that when a rare logical error occurs on the first block, the other blocks are mostly corrupted.  
The analysis in \figref{f:bvsummary} conditions on the first block decoding to~$\ket{-}$, but \tabref{f:singlequerybv} does not place any conditions on it.  
{\bf b},~A similar histogram for the unencoded baseline algorithm shows different behavior.  
Qubit losses are rare, regardless of whether the first qubit is $\ket{-}$ (green), lost (red), or $\ket{+}$~(blue).  
The $\llbracket 4,1,2 \rrbracket$ code turns physical errors into erased code blocks.  
}
\label{f:bv20histogram}
\end{figure}

Computing the success probability conditioned on acceptance makes sense if we treat the BV algorithm as a generic quantum algorithm. 
However, it was originally proposed as a quantum \emph{query} algorithm, in which minimizing the number of oracle queries was key.  
In this light, we can 
ask whether with only one query (trial), the encoded circuit 
beats the unencoded one.  
When the algorithms cannot condition on acceptance, 
it requires guessing~$s$.  With a lost unencoded qubit or an undecodable code block the probability of guessing the corresponding bit of~$s$ is~$1/2$, and the expected Hamming distance 
from~$s$ increases by $1/2$.  
As shown in \tabref{f:singlequerybv}, if the goal is to maximize the probability of guessing~$s$, based on one query to either the unencoded or encoded oracles, then for $n \geq 11$ the unencoded oracle is better.  
On the other hand, if the goal is to minimize the expected Hamming distance to~$s$, then it is better to use the encoded oracle.  
These statements can both hold because the failure modes of the encoded and unencoded algorithms are different (\figref{f:bv20histogram}).  Intuitively, an error on the ancilla qubit in the unencoded algorithm typically results in a string like $1^{n/2}0^{n/2}$, 
contributing $n/2$ to the 
distance from~$s$.  However, with the encoded algorithm 
a faulty ancilla block more likely leads to 
$1^{n/2}\star^{n/2}$, where $\star$ denotes an undecodeable block; 
this contributes only 
$n/4$ to the expected 
distance.  Typical errors in the encoded algorithm are heralded, so they contribute less to the expected 
distance.  

Either quantum oracle is much better than the classical oracle.  
One classical query can only return one bit of~$s$, so the probability of guessing~$s$ is $\Pr[\text{guess}] = 1/2^{n-1}$ and the expected distance 
to~$s$ is $\Ex[\text{Hamming}] = \tfrac{n-1}{2}$.  

\begin{table}
\caption{\label{f:singlequerybv}
Performance of unencoded and encoded Bernstein-Vazirani algorithms with 
one oracle query.  $\Pr[\text{guess}]$ is the probability of guessing the secret~$s$, 
and $\Ex[\text{Hamming}]$ the expected 
distance 
from~$s$.  
For a classical oracle,  $\Pr[\text{guess}] = 1/2^{n-1}$ and $\Ex[\text{Hamming}] = \tfrac{n-1}{2}$.  
}
\setlength{\tabcolsep}{1.9pt}
\begin{tabular}{c@{$\quad$}cc@{$\qquad$}cc}
\Xhline{2\arrayrulewidth}
 & \multicolumn{2}{l}{Unencoded baseline} & \multicolumn{2}{c}{Encoded algorithm} \\
$n$ & $\Pr[\text{guess}]$ & $\Ex[\text{Hamming}]$ & $\Pr[\text{guess}]$ & $\Ex[\text{Hamming}]$ \\
\hline
 7 & 67(4)\% & 0.94 & 75(2)\% & 0.48\\
11 & 69(4)\% & 1.04 & 66(3)\% & 0.90 \\
15 & 61(4)\% & 1.58 & 51(3)\% & 1.45 \\
19 & 49(3)\% & 2.44 & 38(2)\% & 2.29 \\
23 & 35(2)\% & 3.42 & 30(2)\% & 2.97 \\
27 & 28.2(7)\% & 4.12 & 18.5(5)\% & 4.05 \\
\Xhline{2\arrayrulewidth}
\end{tabular}
\end{table}

\section*{Outlook}

In this work, we have created large entangled logical states and implemented 
an algorithm fault tolerantly on encoded qubits, with better-than-physical performance.  
These results highlight the 
features and promise of neutral atoms as a platform for reliable fault-tolerant quantum computing. 
They 
are enabled by the combination of scalability and high-fidelity operations inherent to neutral atoms, with full qubit connectivity. 

All of the experiments 
highlight the usefulness of loss conversion and imaging. 
The encoded experiments demonstrate that logical circuits can correct for and operate effectively with atom loss, which will be a key feature of large-scale neutral atom quantum computers.  

This work, and other recent work with neutral-atom qubits~\cite{BluvsteinHarvard23neutralatoms, quera24neutralatomdistillation}, superconducting qubits~\cite{google24surfacecode} and trapped ion qubits~\cite{Silva24microsoft12qubitcode,ReichardtMicrosoft24tesseract}, highlights a critical advance in the field of quantum computation: the transition from 
noisy, intermediate-scale quantum computations with physical qubits to computational advances achieved with logical qubits.  After a relatively short time from its inception, the neutral atom quantum computing platform has shown significant advancement.  
We expect further advances in 2Q gate fidelities 
and significant scale up towards $\num{10000}$ qubits, including mid-circuit measurement~\cite{norcia2023midcircuit} and continuous atom reloading~\cite{norcia2024iterative}.  
Larger scales will unlock nonlocal error-correcting codes with high encoding rates~\cite{xu2024fastparallelizablelogicalcomputation}. 
The combination of a hardware-optimized qubit virtualization system, complete with 
efficient error correction, with a programmable neutral atom quantum processor 
is a promising path to 
large-depth, logical computation for scientific quantum advantage.

\section*{Acknowledgments}

We thank Zulfi Alam, David Bohn, Jeongwan Haah, Chetan Nayak, Pasi Kostamo, Vadym Kliuchnikov, Dennis Tom, Matthias Troyer, and Jason Zander for their support and feedback during this project.

\bibliographystyle{halpha-abbrv}
\bibliography{bib}

\clearpage
\appendix

\section{Methods}

\subsection{Register loading and atom reuse}\label{loading}

We load atoms into the register using the methods described in \cite{norcia2024iterative}, providing more than enough qubits for the circuit (typically filling 320 register sites to approximately 99\% filling fraction over up to 14 loading cycles).  The register sites are configured in eight pairs of columns, matching the shape of the IZ.  This provides ``highways" through which atoms can be moved with less risk of disturbing other atoms in the register.  By loading more register sites than are used in a given circuit, we can improve data rates by reusing atoms in multiple circuit instances.  To do this, we determine which atoms were lost in a circuit, and refill those sites using extra atoms, then perform the rest of our state preparation protocol (described in the following section).  This can be repeated as long as atoms remain in the register.  

\subsection{State preparation and measurement}\label{selection}

Our platform benefits from the ability to determine successful loading of our desired qubit array before circuit execution, and to separately determine the survival and final state of an atom after the circuit.  We achieve this by performing three images per circuit instance, in addition to images used to initially arrange atoms into the qubit register.  A first ``preselect" image determines whether the atoms were successfully loaded into the register before executing the circuit.  A second ``readout image" comprises the state readout of the circuit.  A third ``survival" image determines whether an atom is still present after the circuit.  In all images, we measure the presence of an atom in state $\ket{1}$.  The pre-select and survival images are combined with optical pumping into $\ket{1}$, so only determine the site occupancy. In the readout image, the optical pumping is omitted and state selectivity is maintained.  Following the preselect image, we perform gray-molasses cooling and optically pump into state $\ket{0}$. 

Aspects of our imaging procedure have been described in \cite{norcia2023midcircuit, norcia2024iterative}. The imaging is performed in a cavity-enhanced 784~nm wavelength 3D lattice by collecting the photons scattered from the cycling transition from $\ket{1}$ to $^3P_1 \ket{3/2, 3/2}$.  Optical pumping and gray-molasses cooling are also performed in the cavity-enhanced lattice.  Between imaging and circuit execution, atoms are transferred between the lattice and register tweezers. 

Typical classification errors, loss probabilities, and spin-flip probabilities for the imaging are provided in~\cite{norcia2024iterative}.  Here, we also characterize state preparation and measurement errors in a manner relevant to our circuit execution by simply preparing either state $\ket{0}$ or $\ket{1}$, and then measuring the state of the atoms, conditioned on the presence of an atom in the survival image.  We obtain combined preparation and readout errors for $\ket 0$ and $\ket 1$ of $1.1(1) \times 10^{-3}$ and $1.6(1) \times 10^{-3}$, respectively.  

\begin{figure}
    \includegraphics[width=1.0\columnwidth]{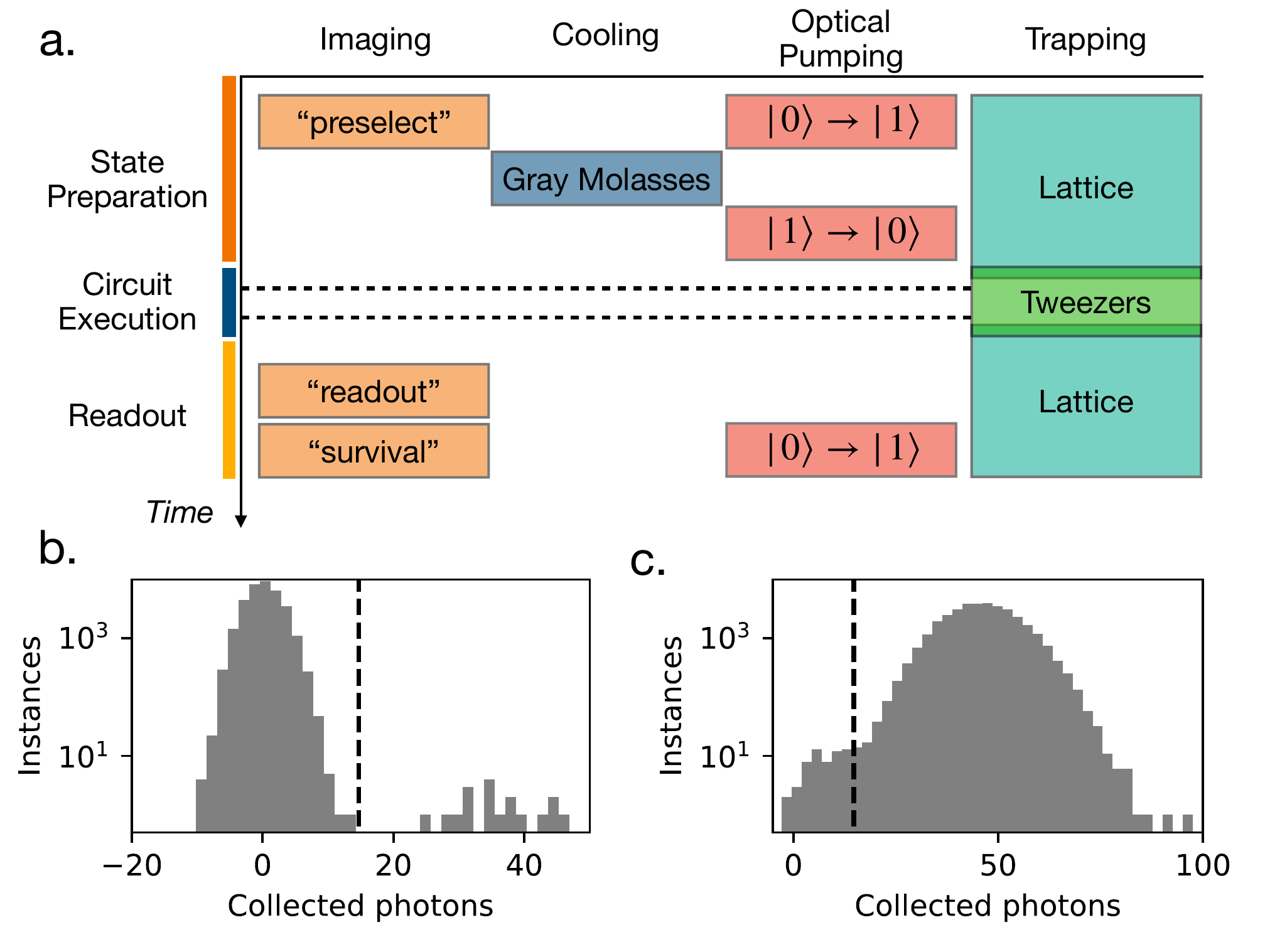}
   \caption{\textbf{a}, Timing diagram of imaging and optical pumping used for circuit execution.  Optical pumping is used both to provide state insensitivity for preselect and survival images, and for state preparation.  \textbf{b}, Photon count histograms recorded after preparing $\ket{0}$ (left) and $\ket{1}$ (right).  The dashed vertical line indicates the count threshold used to distinguish the two states.  Error rates for $\ket{0}$ and $\ket{1}$ are  0.0011(1) and 0.0016(1), respectively.  }
    \label{fig:cavity_SPAM}
\end{figure}

Histograms of the recorded photon counts provide insight into the nature of the errors (\figref{fig:cavity_SPAM}).  For both prepared states, we see clear evidence of a two-part distribution, indicating that the dominant contribution to the error is that the atom actually ends up in the wrong state, rather than overlap of the photon count distribution of dark and bright atoms.  This is likely due to a combination of imperfect optical pumping and spin flips that occur within the image when measuring $\ket{1}$. (Spin flips are primarily due to population of $^3P_1$, so are suppressed when measuring the dark state $\ket{0}$.)  

\subsection{Atom movement} \label{movement}

Arbitrary gate connectivity in our system is obtained by moving selected atoms from the register into an interaction zone (IZ) where two-qubit gates are performed (see \figref{f:system}).  Movement is performed using a set of tweezers generated by a pair of crossed AODs.  Atoms within a single row can be moved in parallel, provided that the move profile does not involve parasitic tweezer spots generated by intermodulation tones crossing the position of the intentional tweezers used for movement.  In practice, we guarantee this condition by requiring parallel moves of more than two atoms to be shape-preserving (stretching and translation of the tweezer locations only).  

Moves are broken up into straight segments, with the movement velocity along each segment following a sinusoidal profile, with a total duration scaled to maintain a fixed peak acceleration (as the square root of the length of the segment).  Move durations are $0.12 \sqrt{L}$~ms, where $L$ is the length of the move in units of the 3.3~$\mu$m grid spacing of the register.  Atoms are handed between static and rearrangement tweezers by ramping the depth of the rearrangement tweezers over 0.4~ms.  For up to 50 moves, we observe a typical loss probability of $5(1) \times 10^{-4}$.  For more moves, loss per move increases.  By performing sideband spectroscopy on the $^1$S$_0$ to $^3$P$_0$ transition after a variable number of moves, we find a heating rate of 0.02(5) quanta along the propagation direction of our clock addressing laser.  This heating rate is too low to explain the increased rate of atom loss above 50 moves, making axial heating a more likely culprit for the excess loss.

In our experiments, atom movement is scheduled using a combination of manual and algorithmic techniques, including the use of graph and annealing algorithms to minimize the number of atom movement operations.

\subsection{Single-qubit gate performance} \label{1Q}

Our single-qubit (1Q) gate scheme is described in detail in Ref.~\cite{muniz2024}, and in that work characterized within a small array.  These gates are performed using two-photon Raman transitions driven by two independently controlled and site-resolved laser beams detuned roughly 10~GHz from the $^3$P$_1$, $F=1/2$ manifold.  Single-qubit operations are performed in parallel on rows of atoms within the register, with the pulse area, phase, and presence of the gates controlled individually on each site.  
We calibrate an $S_X$ gate on each site of the register, and perform virtual $Z$ rotations of arbitrary angle by changing the relative phase of the two drive lasers on subsequent gates.  

\begin{figure}
    \includegraphics[width=1.0\columnwidth]{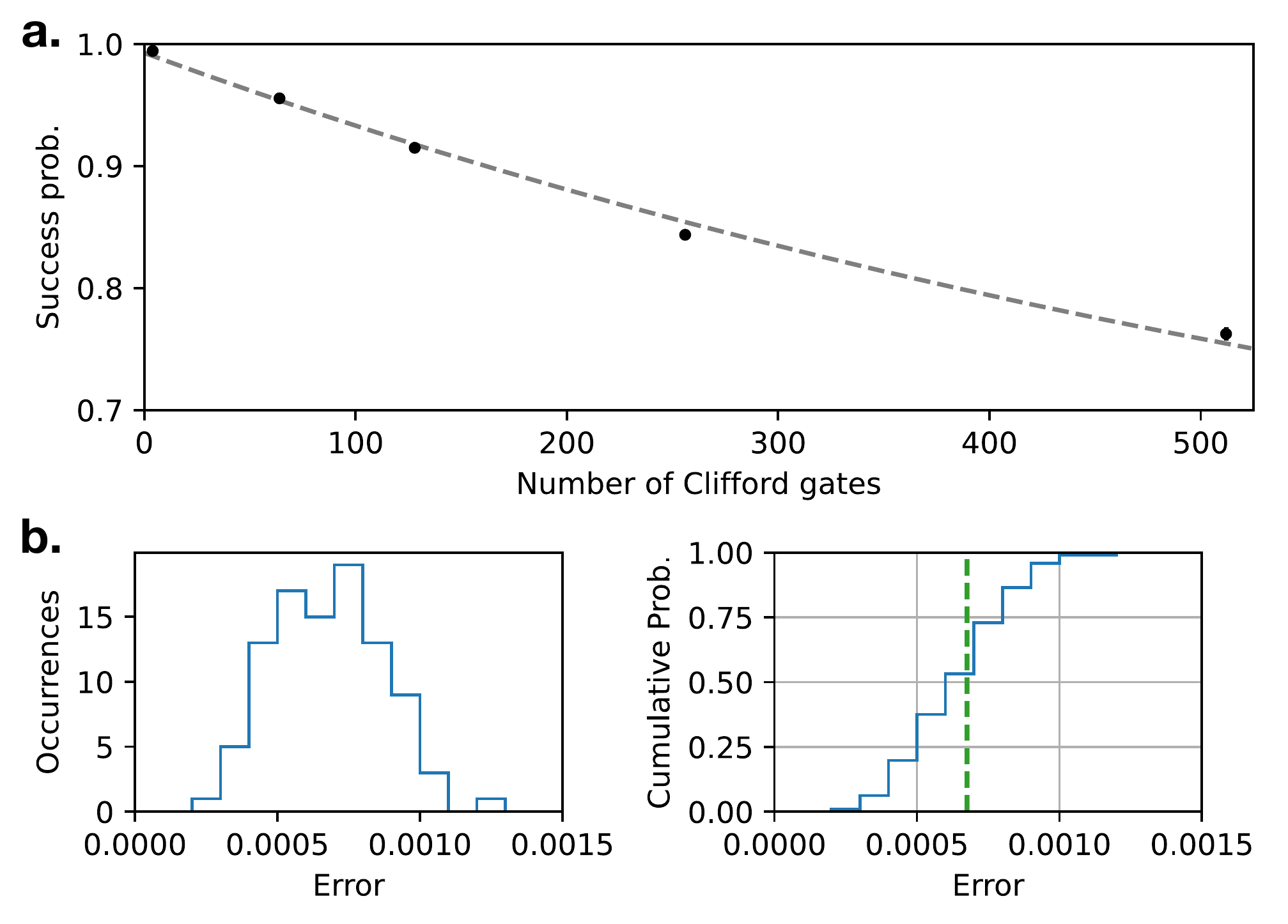}
   \caption{Characterization of single-qubit gates.  \textbf{a}, Array-averaged results of Clifford randomized benchmarking, indicating a per-Clifford-gate error of $6.7 \times 10^{-4}$.  \textbf{b}, Characterization of the spread in per-Clifford-gate errors on different sites within six rows sampled within the register.  }
    \label{fig:1Q}
\end{figure}

Here, we characterize the 1Q gates in the larger register array used in this work. As in \cite{muniz2024}, we perform Clifford randomized benchmarking.  We consider qubits occupying six rows that span those used in our circuits and extract an average error rate per Clifford gate of $6.7 \times 10^{-4}$ (see \figref{fig:1Q}a).  The spread in gate fidelity for different qubit sites is illustrated in \figref{fig:1Q}b.  

In addition to errors occurring during gates, idle errors can also be relevant.  Using Ramsey spectroscopy, we measure a root-mean-squared spread in the qubit frequencies of 0.52~Hz after software compensation of shifts due to magnetic field gradients and average light shifts.  The ensemble-averaged coherence exhibits a Gaussian $1/e$ decay time of 3.5(17) seconds. Uncorrected shifts are likely dominated by inhomogeneous light shifts from the trapping tweezers, which we measure to be typically on the Hz scale for the register tweezers.  Additionally, the rearrangement tweezers lead to shifts on the several-Hz scale, which we calibrate and cancel in software.  These shifts are not fully understood, but may result from finite helicity in the tweezer polarization.  These frequency shifts are small enough that we do not apply dynamical decoupling sequences for the circuits in this work.

\subsection{Two-qubit gate implementation and performance}\label{2Q}

Two-qubit controlled-phase (CZ) gates are performed in the IZ on up to eight pairs of atoms in parallel, using the same scheme described in~\cite{muniz2024}.  
These gates are mediated by excitation to a highly-excited Rydberg state ($65^3\mathrm{S}_1$). 
The sites of the IZ are spaced such that atoms occupying the same pair of sites are in the so-called ``Rydberg blockade" limit, where the strength of interactions between the atoms exceeds other relevant energy scales, while atoms in different pairs interact weakly enough to be treated independently.  
All operations in the IZ are applied in parallel to all atoms within the IZ.  

Atoms are excited to the Rydberg state using a sequential two-photon transition.
State-selectivity is provided by the first stage of the two-photon transition, from $^1$S$_0$, $m_F = -1/2$ to the metastable $^3$P$_0$, $m_F = 1/2$ state.  
Polarization selectivity, along with the narrow linewidth of the transition and slow Rabi rate (10~kHz) relative to the qubit frequency (376~kHz) ensure that only one qubit state is excited to the metastable state.  
From the metastable state, atoms are excited to the Rydberg state using a UV laser pulse (total duration 110~ns) that implements a gate similar to the time-optimal controlled-phase (CZ) gate described in \cite{jandura2022time, pagano2022error, evered2023high}.  While the IZ traps are equally confining for the ground and metastable states, the register traps are near a tune-out wavelength for the metastable states, providing near-zero trapping potential.  Thus, any gate errors that lead to population being left in the metastable state (such as some coherent population errors on either stage of the two-photon transition) are converted to loss and can be detected (see \secref{s:erasure}).  This ``erasure conversion" proves useful in identifying the location of errors \cite{wu2022erasure}.  

We quantify the performance of two-qubit gates using several variations on randomized benchmarking.  To isolate the performance of the gates themselves, without errors from atom movement, we perform interleaved randomized benchmarking (IRB) on static atoms within the IZ.  In this protocol, we compare the return probability after a given number of random two-qubit Clifford operations (plus a return pulse calculated to return to a desired measurement state) to the same circuit with an additional CZ gate interleaved after each Clifford gate (and with an appropriately modified return pulse).  By comparing the probability of returning to the measurement state versus the depth of the Clifford circuit with and without the added CZ gates, we isolate the error associated with the CZ gate alone, averaged over the two-qubit Hilbert space.  Here, the atoms are kept in the IZ for all operations (including 1Q gates, unlike for typical circuits).  Fits to IRB data taken out to depth 34 indicate an infidelity per CZ gate of 0.56(15)\% and additional loss and population leakage outside of the qubit subspace of 0.27(6)\%.

\begin{figure}[htb]
    \includegraphics[width=1.0\columnwidth]{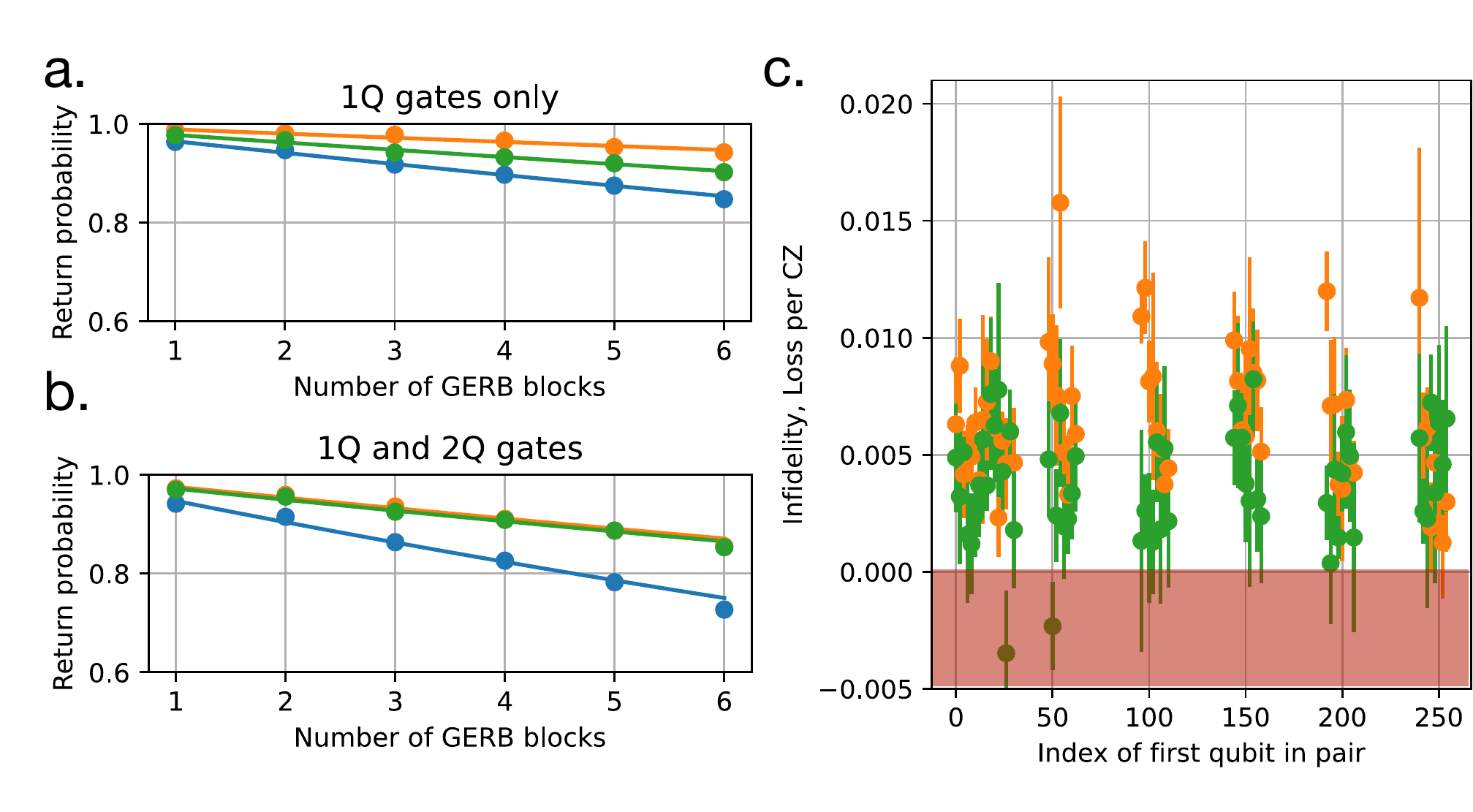}
   \caption{Characterization of CZ gate performance on moved atoms.  \textbf{a}, Echoed randomized benchmarking circuit applied with CZ gates omitted.  Orange, green and blue points represent survival, fidelity for atoms that survive, and combined survival and fidelity, respectively.  Lines represent exponential fits used to extract per-block error and loss rates. \textbf{b}, The same quantities with CZ gates added in (2 CZ gates per block).  \textbf{c}, Measured infidelity (green) and loss (orange) for different qubit origin locations within the register.  Statistical fluctuations may yield negative infidelity estimates because per-block error rates with CZ gates omitted are subtracted from those with CZ gates included.  }
    \label{fig:2Q}
\end{figure}

In actual circuits, atoms are moved between the register and IZ between 1Q operations and 2Q operations.  
Because atom movement can cause heating, which may impact the performance of all following gates, IRB does not cleanly isolate the performance of the CZ gate when atoms are moved between arrays.  For this, we rely on a variant of randomized benchmarking described in~\cite{muniz2024} and similar to that in~\cite{evered2023high}, where concatenated circuit blocks each consist of a random 1Q operation and two CZ gates separated by an echo pulse. A final 1Q gate returns atoms to the measurement basis.  We isolate the contribution from the 2Q gates (along with any loss or infidelity due to associated movements, up to single-qubit phases which are canceled by the echo pulses and characterized separately) by subtracting the per-depth infidelity associated with the 1Q operations.  From this, we extract a per-CZ gate error of 0.39(10)\% and loss of 0.75(10)\%.  We perform this characterization with atoms originating in different rows within the register, as shown in \figref{fig:2Q}, and do not observe a significant dependence on origin row or column.  For comparison, the same procedure applied to atoms that remain static in the IZ for all operations yields a per-gate error rate of 0.35(5)\% and loss of 0.24(6)\%.  

\subsection{Conversion of leakage to loss} \label{s:erasure}

Gate errors in the device, coherent or incoherent, can lead to population and phase errors within the qubit subspace, as well as leakage into other states.  
We operate the machine in a mode that converts population leakage from the qubit subspace into loss, which can then be identified independently from qubit state.  
This is particularly relevant in the context of 2Q gates, where population errors on the $^1$S$_0$ to $^3$P$_0$ or  $^3$P$_0$ to Rydberg state transitions can lead to population leakage.  
Conversion of these errors into loss is achieved by operating our register tweezers at a wavelength near 423~nm where $^3$P$_0$ is nearly untrapped.  (We could choose to depump the $^3$P$_0$ states after each 2Q gate to avoid this loss, but find the conversion of leakage into a detectable error highly desirable.)  

We characterize the effectiveness of the conversion from leakage to loss for $^3$P$_0$ and the Rydberg state by preparing atoms in the register, moving them to the IZ, exciting them to the relevant state, moving them back to the register and then measuring the state and presence of the atoms.  The metastable $^3$P$_2$ states are anti-trapped in our IZ and rearrangement tweezers, and we do not repump these states.  We observe greater than 99\% of atoms to be lost when exciting to the clock state, and roughly 70\% of atoms to be lost when exciting to the Rydberg state.

\section{Four-qubit codes} \label{s:fourqubitcodes}

The $\llbracket 4,2,2 \rrbracket$ code~\cite{VaidmanGoldenbergWiesner96erasure, GrasslBethPellizzari96erasure} is a stabilizer code~\cite{Gottesman97} with stabilizer generators
\begin{equation*}
\begin{array}{cccc}
X&X&X&X\\
Z&Z&Z&Z
\end{array}
\end{equation*}
and logical operators
\begin{equation*}
\begin{array}{rccccc}
\overline{X}_1=&X&X&I&I&\hspace{.31in}{\color{white}.}\\[-.1cm]
\overline{Z}_1=&I&Z&I&Z\\
\overline{X}_2=&X&I&X&I\\[-.1cm]
\overline{Z}_2=&I&I&Z&Z
\end{array}
\end{equation*}
The code can detect any single-qubit error; for example, a single-qubit $Y$ error flips both stabilizer generators.  The code can \emph{correct} for a single-qubit loss.  For example, if the second qubit is lost and this is known (an erasure error), then logical qubit~$1$ can still be decoded using the logical operators $\overline{X}_1 \sim I I X X$, $\overline{Z}_1 \sim ZIZI$, because, having multiplied by a stabilizer, these operators are not supported on the second qubit.  However, once a qubit is lost, the code does not protect against errors on the remaining qubits, e.g., with the second qubit lost an $X_1$ qubit error results in a logical $\overline{X}_1$ error.  

%

Observe that encoded $\ket{00}$ is the four-qubit cat state $\tfrac{1}{\sqrt 2}(\ket{0000}+\ket{1111})$.  As shown in \figref{f:cat24compiledcircuit}, an extra qubit is needed to prepare this state fault tolerantly~\cite{Shor96}.  

\smallskip

The $\llbracket 4,1,2 \rrbracket$ code is actually the same code, but only using one of the two logical qubits.  The other logical qubit is treated as a gauge qubit, meaning we can set it arbitrarily~\cite{Bacon05operator}.  
Unlike the $\llbracket 4,2,2 \rrbracket$ code, the $\llbracket 4,1,2 \rrbracket$ code can sometimes correct two qubit losses.  For example, $\overline X_1$ can still be decoded if both qubits $1$ and $2$ are lost, or both $3$ and $4$, but not any other pair.  

\smallskip

Between two $\llbracket 4,2,2 \rrbracket$ or $\llbracket 4,1,2 \rrbracket$ code blocks, transversal physical CNOT gates fault-tolerantly implement logical transversal CNOT.  
Other logical operations can also be executed fault tolerantly on a $\llbracket 4,2,2 \rrbracket$ code block, including SWAP, CNOT, CZ and Hadamard ($H\otimes H$) on both qubits.
Swapping the middle two physical qubits swaps the logical qubits.  
Swapping the first and third physical qubits implements a logical CNOT gate from~$1$ to~$2$.  Applying the Hadamard gate $H$ transversally implements logical $H\otimes H$ and a logical swap.  
Letting $S_Z = \left(\begin{smallmatrix}1&0\\0&i\end{smallmatrix}\right) = \sqrt{Z}$, and $S_X = \tfrac1{\sqrt 2}\left(\begin{smallmatrix} 1&-i\\-i&1\end{smallmatrix}\right) \propto \sqrt{X}$, observe that 
\begin{align*}
S_Z X S_Z^\dagger &= Y & S_Z Y S_Z^\dagger &= -X & S_Z Z S_Z^\dagger &= Z \\
S_X X S_X^\dagger &= X & S_X Y S_X^\dagger &= Z & S_X Z S_X^\dagger &= -Y.
\end{align*}
Simple stabilizer algebra then shows that applying $S_Z \otimes S_Z^\dagger \otimes S_Z^\dagger \otimes S_Z$ implements a logical CZ gate.
Finally, using $H = S_Z S_X S_Z$, applying $S_X \otimes S_X^\dagger \otimes S_X^\dagger \otimes S_X$ implements a logical dual CZ gate, $(H \otimes H) \,\text{CZ}\, (H \otimes H)$.  
\section{Repeated loss correction and transversal operations for the $\llbracket 4,2,2\rrbracket$~code}

\subsection{Repeated loss correction with CZ logical gates}

The circuits considered so far have only used loss correction and error detection at the very end, when interpreting the final measurements.  
However, it is important to be able to detect and correct errors between the steps of an algorithm, to enable deeper computation.
To that end, we demonstrate repeated fault-tolerant loss correction.  In order to make the circuits more interesting, between rounds of loss correction we apply some of the fault-tolerant logical operations given in \secref{s:fourqubitcodes}.  

Repeated error detection with distance-two codes has been demonstrated on ion traps, including three rounds on eight logical qubits~\cite{self22icebergcode}, and 15 rounds on four logical qubits~\cite{Yamamoto23phaseestimattionerrordetection}.
Demonstrations based on Gottesman's protocol~\cite{Gottesman16smallexperiments} for fault tolerance with the $\llbracket 4,2,2 \rrbracket$ code, but without including repeated error detection, include~\cite{Willsch_2018, Harper19edd,Vuillot_2018,Linke17edexp,Cane2021}.
Repeated error correction, with higher-distance codes, has also been demonstrated on on trapped-ion processors~\cite{honeywell21steane, Postler23steaneec, Silva24microsoft12qubitcode, ReichardtMicrosoft24tesseract} and superconducting processors~\cite{Krinner21repeatedsurfaceec,google23surface,google24surfacecode, amazon24catec}. 

\begin{figure}
\subfigure[]
{\raisebox{0cm}{\includegraphics[scale=1]{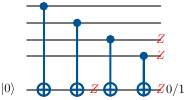}}}
\subfigure[\label{f:xxxxandzzzzsyndrome-no-erasure}]
{\raisebox{0cm}{\includegraphics[scale=1]{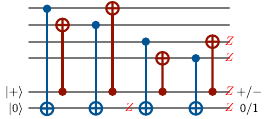}}}
\subfigure[\label{f:xxxxandzzzzsyndromeerasure}]
{\raisebox{0cm}{\includegraphics[scale=1]{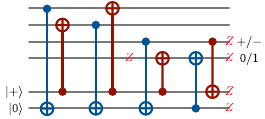}}}
\caption{
{\bf a},
Circuit to measure the parity of four qubits, i.e., the syndrome of $ZZZZ$.  It is not fault tolerant; a single $Z$ fault on the ancilla can spread undetected to a weight-two $Z$ error.  
{\bf b}, The next circuit remedies this, and simultaneously measures $XXXX$, using a trick from~\cite{Reichardt18steane}.  Now the bad $Z$ fault will be detected, or ``flagged," by the $XXXX$ measurement, and similarly the $ZZZZ$ measurement will flag a bad $X$ fault.  
This gives an efficient, fault-tolerant error detection procedure for the $\llbracket 4,2,2 \rrbracket$ code.  However, it does not correct for a lost qubit.  
{\bf c}, The last circuit fixes this, halfway, by switching the direction of the final two CNOT gates~\cite{McEwenBaconGidney23leakage, Chow24leakagedetectionunit}.  
The last two data qubits are replaced with fresh qubits, so in particular if one of them is lost it will be replaced.  Thus all four qubits can be replaced in two rounds of error detection.  
}
\label{f:xxxxandzzzzsyndrome}
\end{figure}

Figure~\ref{f:xxxxandzzzzsyndrome} gives a circuit for fault-tolerant error detection for the $\llbracket 4,2,2 \rrbracket$ code.  It measures the stabilizer generators $X^{\otimes 4}$ and $Z^{\otimes 4}$ in parallel, with one measurement outcome flagging the other for possible correlated errors.  Furthermore, the circuit in \figref{f:xxxxandzzzzsyndromeerasure} replaces two of the data qubits with fresh qubits, so the full data block can be refreshed every two rounds of error detection.  This allows us to use the $\llbracket 4,2,2 \rrbracket$ code to correct a lost qubit during a computation, not just at the very end as in our logical cat state procedure.  

\begin{figure}
\subfigure[]
{\raisebox{0cm}{\includegraphics[scale=.57]{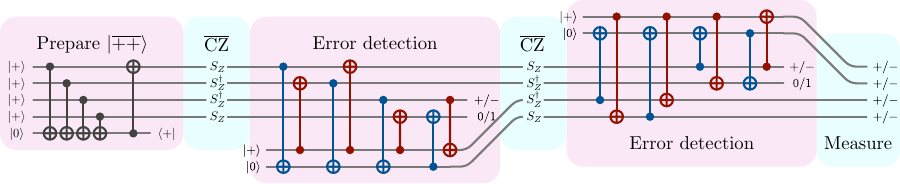}}}
$\;\;$
$\qquad$ 
\subfigure[]
{\raisebox{0cm}{\includegraphics[scale=1]{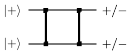}}}
\hspace{2cm}
\subfigure[\label{f:baselineczcz}]
{\raisebox{0cm}{\includegraphics[scale=1]{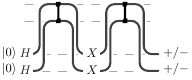}}}
\caption{
(a)~Circuit that applies two rounds of logical CZ gates to $\ket{\overline{++}}$, encoded in the $\llbracket 4,2,2 \rrbracket$ code, with two rounds of error detection  
that refresh all four 
qubits.  
(b)~The corresponding logical circuit.  Since the CZ gates cancel out, both measurements should give~$+$.  
(c)~The baseline circuit run for comparison.  The $X$ gates 
cancel out coherent dephasing errors from the CZ gates and movement.  Both measurements should give $-$.  This baseline performs better than applying the two CZ gates without leaving the interaction zone.  
}
\label{f:repeatedec_czcz}
\end{figure}

We design and execute experiments that repeatedly apply logical CZ gates on a $\llbracket 4,2,2 \rrbracket$ code block.
Recall that $S_Z \otimes S_Z^\dagger \otimes S_Z^\dagger \otimes S_Z$ implements a logical CZ gate.
Error detection with qubit refresh is inserted between each pair of CZ gates, as illustrated in \figref{f:repeatedec_czcz}.
Note that with $j+1$ logical CZ gates, there are $j$ rounds of error detection, with $5 + 8j$ CZ gates applied to $5 + 2j$ qubits.

The encoded experiment can be compared against a corresponding unencoded baseline.  
Given the system architecture, there are a few reasonable baselines from which to choose.  
For example, we may choose to return qubits to the storage zone between each CZ, or not.  
We have chosen to select a baseline in which, as shown in \figref{f:baselineczcz}, between each pair of CZs, both qubits are moved back to the storage zone and Pauli $X$ gates are applied to each of them to cancel coherent dephasing from the movement and two-qubit gates.  This baseline performed the best of those that we tried, {disadvantaging} any comparison between encoded and unencoded.  

\begin{figure}
\includegraphics[scale=.3]{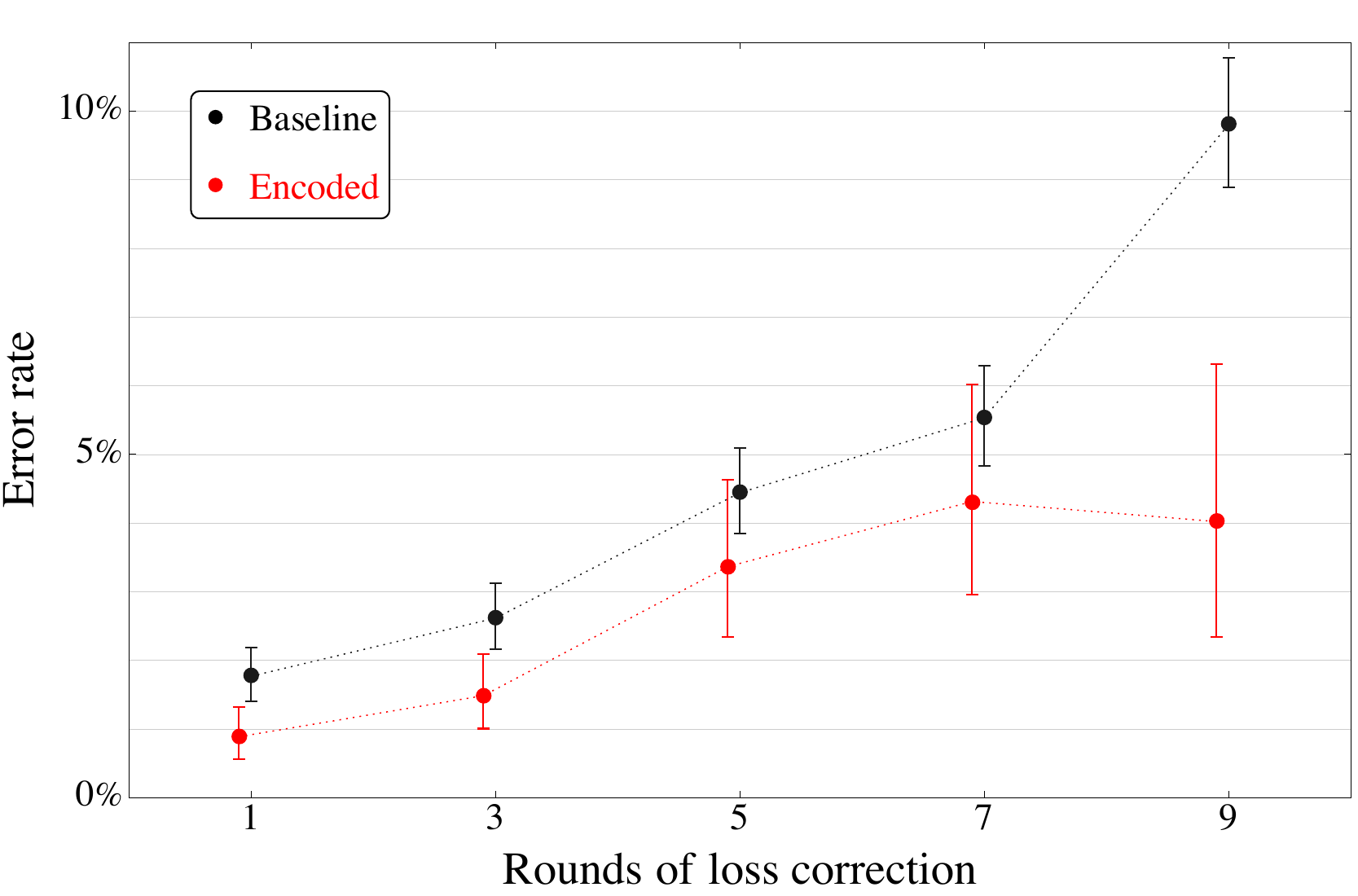}
\caption{
Error rates for $\llbracket 4,2,2 \rrbracket$ code repeated logical CZ experiments, and the corresponding physical baseline, conditioned on no initial loss.  
Error detection between logical gates (\figref{f:repeatedec_czcz}) permits mid-circuit correction of atom loss.
}
\label{f:repeated-cz-plot}
\end{figure}

Results for 
up to 10 logical operations and 9 rounds of loss correction, are shown in \figref{f:repeated-cz-plot}.  
The encoded repeated logical CZ experiments largely beat the corresponding baselines, including, for example, an encoded circuit consisting of 10 logical CZs with 9 loss-correction rounds outperforming an unencoded sequence of 10 CZs. 

\subsection{Random circuits}

\def\CNOT{\mathrm{CNOT}}
\def\CX{\mathrm{CX}}
\def\CZ{\mathrm{CZ}}
\def\SWAP{\mathrm{SWAP}}

As explained in \secref{s:fourqubitcodes}, the $\llbracket 4,2,2 \rrbracket$ code allows a variety of other fault-tolerant logical gates.
Permutations and transversal gates can implement $\CNOT, \CZ, \SWAP$ and $H^{\otimes 2}$, generating a group of size $36$.
To test the performance of these gates, and in the spirit of Ref.~\cite{Gottesman16smallexperiments}, we design experiments by preparing encoded $\ket{00}$ and then applying up to four elements of
the group at random, with error detection between each pair of elements.  
The sequences chosen are, using CNOT = CX for brevity: 
\begin{align} \label{e:randomsequences}
0:&& &H^{\otimes 2} \, \CX \\
1:&& &I,\; \CX \, \CZ \, H^{\otimes 2} \, \CX \nonumber\\
2:&& &
H^{\otimes 2} \, \SWAP \, \CX,\;
\CZ \, H^{\otimes 2} \, \SWAP,\; 
H^{\otimes 2} \, \SWAP \, \CZ \nonumber\\
3:&& &
\CX \, H^{\otimes 2} \, \CX,\;
H^{\otimes 2},\;
\SWAP \, \CX,\;
\CZ \, \CX \, H^{\otimes 2} \nonumber
\end{align}
The encoding of sequence $j$ involves $j$ rounds of error detection.
Note that the error detection circuits used here are those of~\figref{f:xxxxandzzzzsyndrome-no-erasure}, instead of those of~\figref{f:xxxxandzzzzsyndromeerasure}; therefore they correct for lost qubits only at the end.  

The choice to error-detect between group elements is justified by imagining a computation with many $\llbracket 4,2,2 \rrbracket$ blocks.  In that scenario, it is natural to write a circuit as alternating layers of CNOTs between blocks, and group elements within each block. It is then also natural to schedule error detection following each pair of layers in order to control the spread of errors among blocks.

\begin{figure}
\includegraphics[scale=.3]{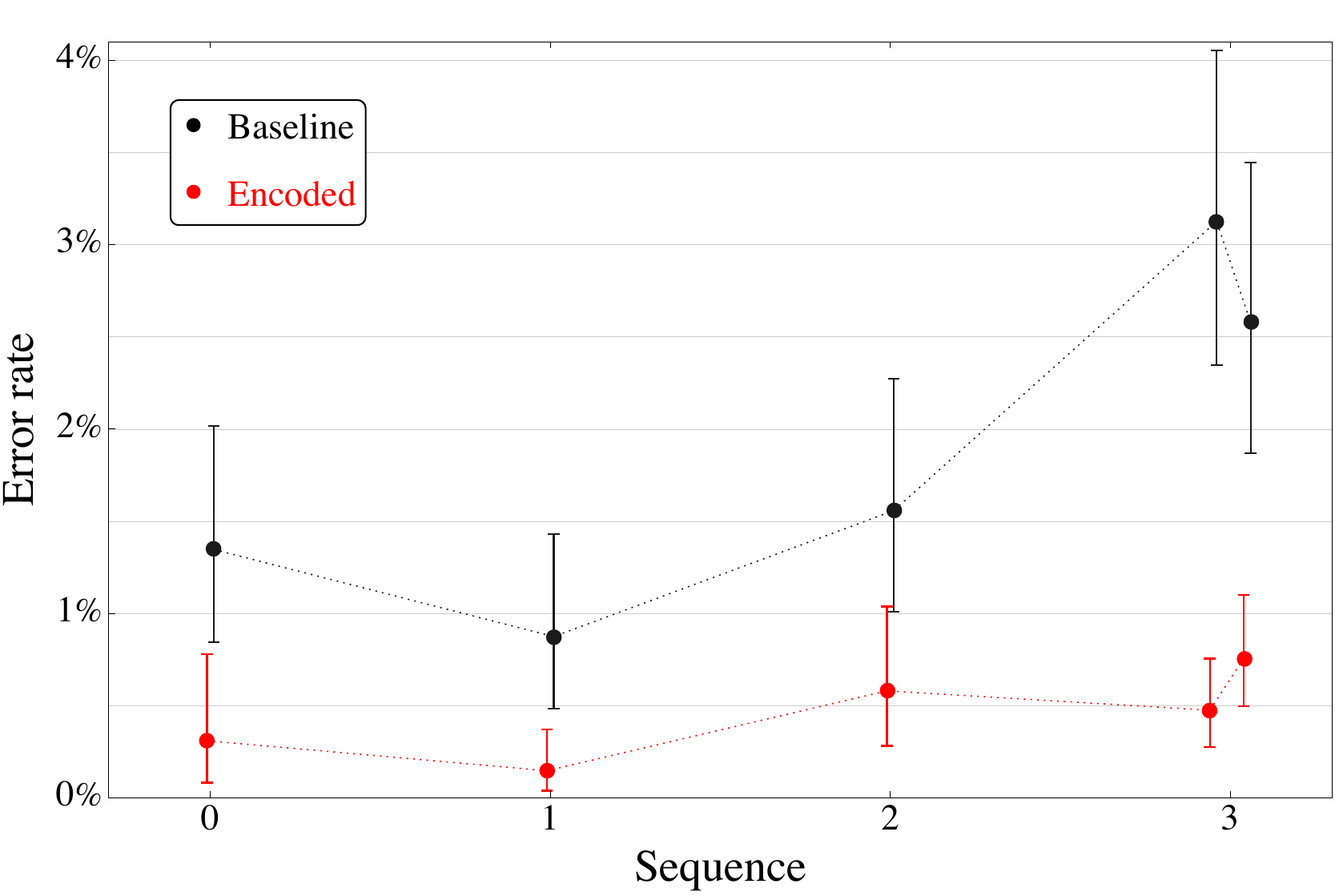}
\caption{
Error rates for $\llbracket 4,2,2\rrbracket$ code random logical circuits, conditioned on no initial loss.  
Sequence labels are from the sequences given in Eq~\eqnref{e:randomsequences}.
}
\label{f:422-random-plot}
\end{figure}

The output of each random sequence is a stabilizer state with two generators.  
Each generator is measured to determine success of failure of the experiment.
The encoded circuits can be directly compared to corresponding physical baselines that execute the same gate sequence.
As shown in \figref{f:422-random-plot}, the encoded circuits outperform the corresponding physical baselines.

\section{State preparation with the $\llbracket 16,6,4 \rrbracket$~code}

\begin{figure}
\begin{tabular}{c@{\quad}c}
\hspace{-.4cm}
\subfigure[\label{f:tesseractft}]{\raisebox{.2cm}
{\includegraphics[scale=.4]{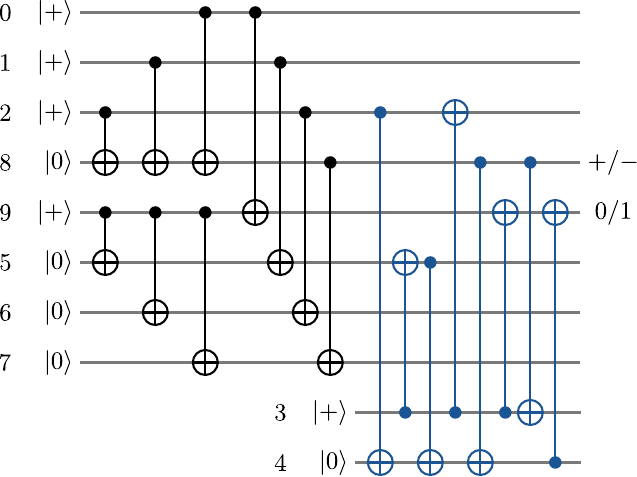}}}
&
\subfigure[\label{f:tesseractnft}]{\raisebox{.2cm}
{\includegraphics[scale=.4]{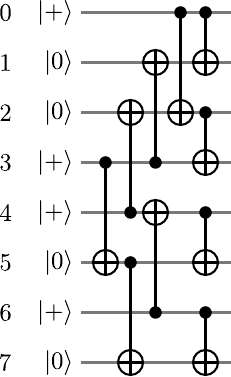}}}
\end{tabular}
\begin{tabular}{c@{\qquad}c}
\hspace{-.4cm}
\subfigure[\label{f:tesseract250}]{\raisebox{.2cm}
{\includegraphics[scale=.4]{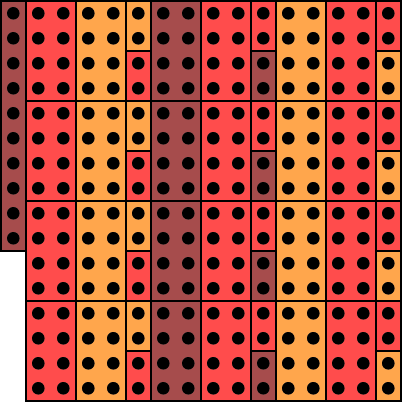}}}
&
\subfigure[\label{f:tesseract256}]{\raisebox{.2cm}
{\includegraphics[scale=.4]{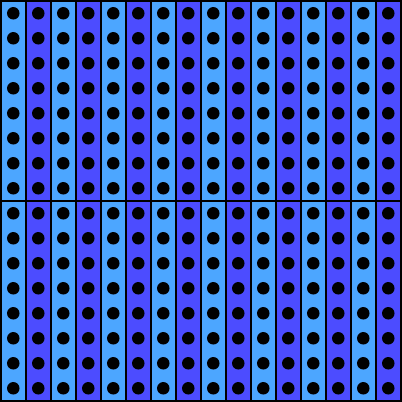}}}
\end{tabular}
\hspace{-.4cm}
\subfigure[\label{f:tesseractbar}]{\raisebox{0cm}
{\includegraphics[scale=.5]{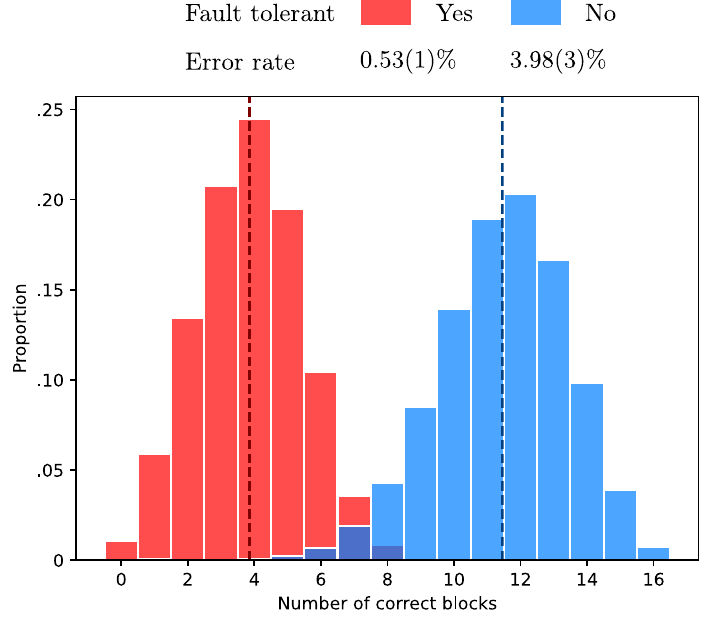}}}
\caption{
{\bf a}, Fault-tolerant preparation of encoded $|{+}\rangle^{\otimes3}$ in the $\llbracket 8,3,2 \rrbracket$ code.
The blue part uses the gadget in \figref{f:xxxxandzzzzsyndromeerasure} to verify two of the stabilizers and swap qubits 3, 4 with 8, 9.
The prepared state on qubits 0 to 7 is accepted only if qubits 8 and 9 are present and have trivial measurement outcomes.
{\bf b}, Non--fault-tolerant preparation.
{\bf c}, Layout of 250 atoms for preparing 25 copies of encoded $|{+}\rangle^{\otimes3}$ using {\bf a}.
Colors only indicate the partition.
Each $4\times 2$ tile joins a nearest same colored $2\times 1$ tile.
{\bf d}, Layout of 256 atoms for preparing 32 copies of encoded $|{+}\rangle^{\otimes3}$ using {\bf b}.
{\bf e}, Histogram of number of correct blocks of encoded $|{+}0{+}0{+}0\rangle$ in the $\llbracket 16,6,4 \rrbracket$ code per shot, including shots with both $X$ and $Z$ basis measurements.
Vertical lines indicate averages of 3.85(1) and 11.44(3) for the fault-tolerant and non--fault-tolerant protocol, respectively.}
\end{figure}

We use all the 256 atoms to prepare error-corrected logical qubits encoded in blocks of the $\llbracket 16,6,4 \rrbracket$ tesseract code~\cite{ReichardtMicrosoft24tesseract}.
To this end, we first prepare multiple copies of $|{+}\rangle ^{\otimes 3}$ encoded in the $\llbracket 8,3,2 \rrbracket$ color code~\cite{Campbell16eight32colorcode}, and then randomly pair them up, each pair being a copy of $|{+}0{+}0{+}0\rangle$ encoded in the tesseract code~\cite{ReichardtMicrosoft24tesseract}.

We consider both fault-tolerant and non--fault-tolerant preparations of encoded $|{+}\rangle ^{\otimes 3}$ in the $\llbracket 8,3,2 \rrbracket$ code, as in \figref{f:tesseractft} and \figref{f:tesseractnft}, respectively.
Compared to~\figref{f:tesseractnft}, \figref{f:tesseractft} uses two extra qubits to measure the values of two stabilizers of the encoded $|{+}\rangle ^{\otimes 3}$, and rejects upon any nontrivial measurement outcome or detected loss.
For the non--fault-tolerant protocol, we first prepare 32 copies of encoded $|{+}\rangle ^{\otimes 3}$, using the layout in \figref{f:tesseract256}, and then randomly pair them up into 16 copies of encoded $|{+}0{+}0{+}0\rangle$.
For the fault-tolerant protocol, we instead prepare 25 copies of encoded $|{+}\rangle ^{\otimes 3}$ (with six atoms unused), as shown in \figref{f:tesseract250}, and only pair up accepted ones.

In both protocols, we measure all the atoms of the prepared tesseract code blocks in the same $X$ or $Z$ basis, and then decode.
Specifically, an encoded $|{+}0{+}0{+}0 \rangle$ block can be either ``rejected'' due to $\ge2$ errors or losses detected, or ``correct'' if all the three deterministic values of logical $X$ or $Z$ operators are trivial, or ``wrong'' otherwise.

For the fault-tolerant protocol, we run 1980 and 2000 trials for $X$ and $Z$ basis, respectively; for the non--fault-tolerant protocol, we run 960 and 1000 trials, respectively.
For each of these trials, we repeat the random pairing of the (accepted) blocks of encoded $|+\rangle ^{\otimes 3}$ independently for 100 shots.
Results of these shots are shown in \figref{f:tesseractbar}.
Specifically, we calculate the error rate, defined as the ratio of the number of wrong blocks over the number of correct or wrong blocks, both numbers summing over all shots of $X$ and $Z$ bases.
We observe an error rate of $0.53(1)\%$ for the fault-tolerant protocol, nearly one order of magnitude smaller than $3.98(3)\%$ for the non--fault-tolerant protocol.
However, the average number of correct blocks per shot is 3.85(1) for the fault-tolerant protocol, compared to 11.44(3) for the non--fault-tolerant protocol.
This demonstrates a tradeoff between error rate and acceptance rate enabled by loss conversion.


\ifx\compilefullpaper\undefined  
\end{document}
\fi